\documentclass[amsmath,amssymb,aps,letterpaper,prb,twocolumn,longbibliography,superscriptaddress]{revtex4-2}
\usepackage{tikz}
\usepackage[english]{babel}
\usepackage[utf8]{inputenc}
\usepackage[T1]{fontenc}
\usepackage{indentfirst}
\usepackage{amsmath}
\usepackage{amssymb}
\usepackage{amsthm}
\usepackage{proof}
\usepackage{eufrak}
\usepackage{graphicx}
\usepackage{psfrag}

\allowhyphens

\newcommand{\La}{\mathcal{L}}

\newcommand{\HH}{\mathcal{H}}
\newcommand{\complex}{\mathbb{C}}

\newcommand{\valos}{\mathbb{R}}
\newcommand{\eps}{\varepsilon}

\newtheorem{defdef}{Definition}

\newcommand{\ket}[1]{{\left|#1\right\rangle}}
\newcommand{\bra}[1]{{\left\langle #1\right|}}


\usepackage{ifpdf}

\ifpdf
\usepackage{epstopdf}
\usepackage{float}
\usepackage[pdftex,colorlinks,urlcolor=blue,citecolor=blue,linkcolor=blue]{hyperref}
\else
\usepackage[hypertex,colorlinks,urlcolor=blue,citecolor=blue,linkcolor=blue]{hyperref}
\fi
\begin{document}
\numberwithin{equation}{section}

\title{Weak ergodicity breaking with isolated integrable sectors}
\author{Hosho Katsura}
\affiliation{Department of Physics, University of Tokyo, 7-3-1 Hongo, Bunkyo-ku, Tokyo 113-0033, Japan}
\affiliation{Institute for Physics of Intelligence, University of Tokyo, 7-3-1 Hongo, Bunkyo-ku, Tokyo 113-0033, Japan}
\affiliation{Trans-scale Quantum Science Institute, University of Tokyo, Bunkyo-ku, Tokyo 113-0033, Japan}
\author{Chihiro Matsui}
\affiliation{Department of Mathematical Sciences, University of Tokyo, 3-8-1 Komaba, Meguro-ku, Tokyo 153-8914, Japan}
\author{Chiara Paletta}
\affiliation{ Department of Physics, Faculty of Mathematics and Physics, University of Ljubljana, Jadranska 19, SI-1000 Ljubljana, Slovenia}
\author{Bal\'azs Pozsgay}
\affiliation{MTA-ELTE “Momentum” Integrable Quantum Dynamics Research Group, Department of Theoretical Physics, ELTE Eötvös
  Loránd University, P\'azm\'any P\'eter stny. 1A, Budapest 1117, Hungary}

\begin{abstract}
  We consider spin chain models with local Hamiltonians that display weak ergodicity breaking. In 
  these models, the majority of the eigenstates  
  are thermal, but there is a distinguished subspace of the Hilbert space in which ergodicity is broken. We achieve such a weak breaking by embedding selected integrable models into larger Hilbert spaces of otherwise chaotic models. 
The integrable subspaces do not have a tensor product structure with respect to any spatial bipartition, therefore our constructions differ from certain trivial embeddings. 
We consider multiple mechanisms for such an embedding, and we also review previous examples in the literature. Curiously, all our examples can be seen as perturbations of models with Hilbert space fragmentation, such that the perturbed models are not fragmented anymore.
\end{abstract}

\maketitle

\section{Introduction}

Ergodicity is a central concept in theoretical physics, that is essential for the foundations of statistical mechanics,
both in the classical and the quantum realms. One of the interesting questions that received considerable attention is:
What are the possible ways of ergodicity breaking, and what are their experimental consequences?

The precise definitions of ergodicity might depend on the context. In quantum many-body physics, the most often used
definition is that a system is ergodic if it thermalizes. This means that in typical nonequilibrium situations, the
time evolution of the isolated system is such that the emerging steady states are locally indistinguishable from thermal
ensembles. It is now understood that thermalization happens on the level of eigenstates. More precisely,  the so-called
Eigenstate Thermalization Hypothesis (ETH) states that the reduced density matrices of excited states of a large many-body system are typically indistinguishable from the reduced density matrices computed from thermal ensembles (for a review,
see \cite{eth-review}). This implies that ergodicity can be broken if the ETH is not satisfied.

The various mechanisms of ergodicity breaking can be grouped into two large categories \cite{moudgalya2019thermalization}. We say that a model has strong
ergodicity breaking if a typical eigenstate breaks the ETH. In contrast, a model has weak ergodicity breaking if the
number of eigenstates not satisfying ETH becomes negligible when compared to the dimension of the Hilbert space in the
thermodynamic limit.

The distinction between strong and weak ergodicity breaking has immediate consequences in
experimental situations. In the case of strong breaking, the non-equilibrium processes will typically demonstrate the lack of thermalization. In contrast, systems with weak ergodicity breaking will appear ergodic in almost all
non-equilibrium processes, but in selected scenarios, they will display the departure from thermalization. A famous
example for such behavior was the observation of long-lived oscillations in a Rydberg atomic system
\cite{scars-nature}. This experiment led to the discovery of the so-called quantum many-body scars \cite{scars-elso, PXP-2}, which are exceptional eigenstates in the middle of the spectrum that break the ETH. Weak ergodicity breaking and quantum many-body scars are reviewed, for example, in \cite{fragmentation-scars-review-2,hfrag-review}.

In this work, we provide a systematic analysis of a special way of {realizing} weak ergodicity breaking: we focus on the models that have an isolated integrable sector. This means that most of the eigenstates of the model are thermal, but there is a special sector of the full Hilbert space that realizes an integrable model. We focus on cases where the integrable sector grows exponentially with the length of the spin chain; nevertheless its dimension becomes negligible when compared to the dimension of the full Hilbert space. Such models are clearly distinct from models supporting quantum many-body scars: in those cases, there is a ``scarred'' subspace, which grows at most polynomially with the system size.

Our models are different from many other proposals for embedding integrable models into larger Hilbert spaces. Most of the embeddings that appeared in the literature have strong ergodicity breaking, which means that the complement of the integrable subspace is not thermalizing either; this will be discussed in Sec. \ref{sec:pre}. Alternatively, the embedding might be trivial in the sense that the image of the embedding takes a tensor product structure with respect to any spatial bipartition; such models are reviewed in Sections \ref{sec:trivial}.

To our best knowledge, the only non-trivial example in the literature for weak ergodicity breaking by an integrable subspace is the so-called XX spin ladder model treated in \cite{znidaric-coexistence}. That model is not integrable, most eigenstates are thermal and the heat transport is found to be diffusive. Nevertheless, there are isolated integrable sectors which support ballistic transport, and the embedding is not trivial. The key property of the model is that certain spin patterns are preserved in a properly selected basis, and states with the distinguished patterns form the basis for the integrable subspace.

In this work we find similar behaviour also in other examples. Moreover, 
we find that weak ergodicity breaking by integrable subspaces is closely related to Hilbert space
fragmentation. All the examples that we find can be seen as perturbations of models with Hilbert space fragmentation,
and this includes the model of \cite{znidaric-coexistence}. Fragmentation means that the full Hilbert space is split
into exponentially many disconnected sectors, and as an effect, the models strongly break ergodicity
\cite{fragment-fracton-2,tibor-fragment,hfrag-review,fragmentation-scars-review-2}. 

It is known that fragmented models often have a few integrable sectors, for examples see \cite{tibor-fragment,fragm-commutant-1,moudgalya2019thermalization,folded0b,xavier2021fractons,dhar-alternative-to-folded}.
Such models appeared even much earlier in the literature, although the term ``Hilbert space fragmentation'' did not exist at that time, for examples see \cite{nicolai1976supersymmetry,batista1995superconductivity,zhang1997charge,tJz}.  A more recent example of a fragmented model with an integrable sector was treated recently in \cite{chihiro-embedding}, although the fragmentation was not stressed there. We treat the model of \cite{chihiro-embedding} in detail in Sec. \ref{sec:MM}.

In some cases, a fragmented model might be integrable in all sectors. Perhaps the most important examples are the so-called folded XXZ model \cite{folded0a,folded0b,folded1,folded2,sajat-folded} and the Temperley-Lieb models of \cite{read-saleur}. The so-called Maassarani-Mathieu spin chain \cite{su3-xx} and its generalizations such as the so-called XXC models and the multiplicity models of Maassarani \cite{XXC,multiplicity-models} are also integrable and fragmented, and they are related to the folded XXZ model \cite{sajat-folded,sajat-hardrod,folded-XXZ-stochastic-1}.  

The key observation of our work is that if one takes a fragmented model with at least one integrable sector, then very
often one can add perturbations that undo the fragmentation in the majority of the Hilbert space, keeping the integrable
sector invariant. Perhaps surprisingly, even the previously published model of \cite{znidaric-coexistence} fits into
this framework, as we show below in Sec. \ref{sec:ladder}. 

We should note that a different type of weak ergodicity breaking by integrable sectors was studied recently in \cite{mussardo-embedding}. That work differs from ours because there the dimension of the integrable sector grows only polynomially with the volume if one keeps the dimension of the local spaces fixed. We treat models where the integrable sector grows exponentially with the volume.

Our paper is structured as follows. In Sec. \ref{sec:pre}, we provide all the definitions and basic facts needed for our discussion. In Sec. \ref{sec:trivial}, we discuss trivial embeddings of integrable models. In Sections \ref{sec:MM}-\ref{sec:constr}, we discuss various examples of different mechanisms for weak ergodicity breaking with embedded integrable sectors. In Sec. \ref{quantumcircuits}, we consider the embeddings of integrable models in certain types of quantum circuits.
Section \ref{sec:concl} includes our Conclusions.

\section{Preliminaries}

\label{sec:pre}

We consider spin chain models with local spaces $\complex^D$ with some $D\ge 2$. In a finite volume $L$, the Hilbert space is $\HH=\otimes_{j=1}^L \complex^D$.

The Hamiltonian is an extensive operator with a local operator density:
\begin{equation}
    H=\sum_j h(j).
\end{equation}
Here $h(j)$ is assumed to act on a few sites around site $j$. In most of our examples, we treat translationally invariant models.
One of our examples is the spin ladder model of \cite{znidaric-coexistence}, which can be regarded as translationally invariant with an elementary cell consisting of two sites.

We consider both periodic and open boundary conditions.

\subsection{Ergodicity and its breaking}

Ergodicity can be defined in multiple ways. We say that a model is ergodic if it satisfies the Eigenstate Thermalization
Hypothesis (ETH).
The ETH is regarded as the main mechanism that guarantees thermalization of isolated quantum systems in non-equilibrium
situations \cite{eth1,eth2,rigol-eth,eisert-thermalization-eth-review}.

\begin{defdef}
  A model satisfies the ETH if
  in the thermodynamic limit the  expectation values $\bra{\psi}A\ket{\psi}$ of local operators $A$ in excited
    states $\ket{\psi}$ are smooth 
    functions of the overall energy density of the state, and do not depend on other details of the state $\ket{\psi}$
 \end{defdef}

The ETH implies that reduced density matrices in excited states become identical to those of thermal ensembles,
thus the individual states become indistinguishable from thermal averages. Thermalization happens on the level of
eigenstates. 

There are various ways of breaking ergodicity, and the various mechanisms can be grouped into two families: strong and
weak breaking of ergodicity. To introduce these families, let us consider a model in which some of the eigenstates
$\ket{\psi}$ are such that the reduced density matrices do not reproduce the thermal averages. Let us denote the
subspace spanned by these states as $\mathcal{H}_{\rm non-ETH}$. 

\begin{defdef}
We say that a system is weakly breaking ergodicity, if the ratio of the dimensions of $\mathcal{H}_{\mathrm{non-ETH}}$ and that of the full Hilbert space approaches zero in the
thermodynamic limit. A system is strongly breaking ergodicity, if the dimension of $\mathcal{H}_{\rm non-ETH}$ remains comparable to
the total dimension of the Hilbert space.
\label{defweakergb}
\end{defdef}

There are three main mechanisms of strong ergodicity breaking: Integrability, many-body localization, and (strong) Hilbert space
fragmentation. Below we will discuss the concepts of integrability and Hilbert space fragmentation; many-body localization is not treated in this work.

Weak ergodicity breaking is reviewed in \cite{fragmentation-scars-review-2}. Its two main mechanisms are quantum many-body scarring and weak Hilbert space fragmentation. Our work provides another mechanism, namely, by embedding integrable models into isolated subspaces of an otherwise chaotic model.

\subsection{Strong ergodicity breaking: Integrability}

There is no all encompassing definition of integrability in the case of quantum integrable models \cite{caux-integrability}. For our purposes, the following definition is convenient: 

\begin{defdef}
We say that a spin chain Hamiltonian with a local operator density is integrable, if there exist a family $\{Q_\alpha\}$ of operators such that
\begin{itemize}
    \item every $Q_\alpha$ is an extensive operator with a local operator density
\item     the operators form a commuting family:
\begin{equation}
[Q_\alpha,Q_\beta]=0, \,\,\,{\forall \, \alpha, \beta}
\end{equation}
\item the Hamiltonian is a member of the family
\item the number of the operators that can be defined in a finite volume $L$ grows at least linearly with $L$.
\end{itemize}
\end{defdef}

\subsection{Hilbert space fragmentation}

Hilbert space fragmentation also leads to ergodicity breaking \cite{tibor-fragment,fragment-fracton-2}, see the reviews \cite{hfrag-review,fragmentation-scars-review-2}. It can be defined formally via the so-called commutant algebra \cite{fragm-commutant-1}, but here we adopt a definition that is more direct, although less precise.

\begin{defdef}
    We say that a spin chain model has Hilbert space fragmentation if its finite volume Hilbert space splits into sectors  $\HH=\oplus_{k=1}^{N_H} \HH_k$ such that
\begin{itemize}
\item the Hamiltonian has no transition matrix elements between different sectors
\item the number $N_H$ of the sectors grows exponentially with the volume $L$
\item the sectors can be constructed without having to find the exact eigenstates of the Hamiltonian. 
\end{itemize}    
\end{defdef}
The reason for adding the last condition is the following. If one has an exact knowledge of all the eigenstates
$\ket{n}$ of the model in every finite volume $L$, then the individual eigenstates could be viewed as separate sectors,
thereby satisfying the first two requirements. However, the key property of Hilbert space fragmentation is that there
are certain simple kinetic or other types of constraints, with which the fragments can be constructed in a straightforward way, without needing to know the exact eigenstates.

The fragmented sectors are often obtained via the Krylov basis. The idea is to take a certain initial vector
$\ket{\Psi_0}$, which might be a product state or some other locally constructed state, and to construct the set of
vectors 
\begin{equation}
    \{H^k\ket{\Psi_0}, k=0, 1, \dots, n\},
\end{equation}
where $n$ is chosen as the smallest integer such that $H^{n+1}\ket{\Psi_0}$  is linearly dependent on the set of
vectors. In generic ergodic models, the Krylov basis typically spans the full Hilbert space. If there are global
symmetries present, then the Krylov basis is expected to span the full sector with a given set of quantum numbers, if
the initial state is an eigenvector of the symmetries. In contrast, if the Hilbert space is fragmented, then the Krlylov spaces are typically exponentially smaller.

There are two main types of Hilbert space fragmentation:
\begin{defdef}
A model has strong Hilbert space fragmentation if the ratio of the dimensions of the largest sector and the full Hilbert space (or the sub-sector allowed by global symmetries that contains the largest sector) vanishes in the thermodynamic limit.
    A model has weak Hilbert space fragmentation, if the dimension of the largest sector of the full Hilbert space (or the sub-sector allowed by global symmetries) remains comparable with the full Hilbert space.
\end{defdef}

If a model has Hilbert space fragmentation, then the total symmetry algebra associated with the Hamiltonian is exponentially large in the volume \cite{fragm-commutant-1}. This can be understood intuitively as follows. If the Hilbert space splits into
exponentially many disconnected sectors, then we can obtain an exponentially large symmetry algebra simply by constructing the projectors associated with the different sectors. All of these projectors commute with the Hamiltonian by definition, and they belong to the commutant algebra.

Typically these projectors are nonlocal operators, but the symmetry algebra can often be generated by Matrix Product Operators (MPO's) \cite{fragm-commutant-1}. An MPO symmetry can be seen as a simple generalization of a strictly local symmetry operation. An MPO has finite entanglement entropy  (in operator space) if its bond dimension is finite.

Instead of constructing the projectors themselves, in many cases, it is possible to extract a certain type of nonlocal
classical information from the sectors and this classical information can be used as a label for the individual
sectors. Such classical information is always conserved during time evolution because they label the sectors. In
earlier literature, such pieces of classical conserved quantities were called ``irreducible string''
\cite{trimer1,trimer2,folded-XXZ-stochastic-1,dhar-symmetries,dhar-alternative-to-folded} (see also \cite{dhar-alternative-to-folded}). These
earlier works dealt with stochastic models, but the formalism and the mathematical properties of fragmentation are the
same as in the quantum spin chains. In recent literature, the conserved quantities in question were called
``Statistically localized integrals of motion'' \cite{tibor-fragm-SLIOM}. In this work, we will use the term
``irreducible string''. 

\subsection{Strong ergodicity breaking with integrable subspaces}

In this work, we will show a mechanism whereby ergodicity can be weakly broken by an isolated integrable
subspace. However, before turning to our results, it is useful to discuss existing work on Hilbert space fragmentation
with integrable subspaces. Multiple works considered fragmented models which are chaotic in almost all sectors, except for a few integrable sectors. Due to the fragmentation, these models are strongly breaking ergodicity.

Now we list a few such models. We do not claim that this list is complete.
\begin{itemize}
\item The dipole conserving spin-$1$ model of \cite{tibor-fragment}. It has at least one integrable subspace, which realizes the XX model. 
\item The pair hopping spin-$1/2$ model of \cite{moudgalya2019thermalization}. It has multiple integrable subspaces, all of which can be mapped into the spin 1/2 XX model of different system sizes. 
\item The model of \cite{chihiro-embedding}, which embeds the integrable XXZ model (with arbitrary anisotropy $\Delta$) into a strongly fragmented spin-$1$ model.
\end{itemize}

\section{Weak ergodicity breaking by trivial embeddings}

\label{sec:trivial}

In this section, we treat trivial examples 
exhibiting weak ergodicity breaking with isolated integrable subspaces. First, we
consider a concrete model in Subsection \ref{sec:trivconcr}, and 
then provide a more general treatment in
\ref{sec:trivgen}.

\subsection{A spin-$1$ chain}

\label{sec:trivconcr}

We consider a spin-$1$ chain, which  belongs the family of models treated in \cite{sato1996first,sato-embedding}.

In order to write down the Hamiltonian, we introduce a few notations. We choose the local basis states as $\ket{a}$ with
$a=0,1,2$. The state $\ket{0}$ is interpreted as the vacuum, and the states $\ket{1}$ and $\ket{2}$ are seen as
excitations with an internal degree of freedom. This internal degree of freedom can be called ``color'' or ``spin''. In
this work, we will use the term ``color''.

For a local Hilbert space we introduce the creation/annihilation operators of particles of color $a$
\begin{equation}
    \sigma^{+,(a)}=\ket{a}\bra{0},\quad 
   \sigma^{-,(a)}=\ket{0}\bra{a},  \quad a=1,2
\end{equation}
and also the particle number operators
\begin{equation}
    N^{(a)}=\sigma^{+,(a)}\sigma^{-,(a)}=\ket{a}\bra{a},\quad a=1,2
    \label{eq:Naop}
  \end{equation}
  together with the combined particle number operator 
  \begin{equation}
    N^{\rm tot}=N^{(1)}+N^{(2)}.
  \end{equation}
We will also use the local projection operator onto the vacuum state defined as
\begin{equation}
    P=1-N^{\rm tot}=\ket{0}\bra{0},
    \label{eq:Pop}
\end{equation}
where $1$ is the identity operator 
and the local ``color raising and lowering operators'' are defined now as
\begin{equation}
  S^+=\sigma^{+,(2)}\sigma^{-,(1)},\qquad  S^-=\sigma^{+,(1)}\sigma^{-,(2)} .
\end{equation}
Below these operators will get an index corresponding to the site where they act.

Then the Hamiltonian is written as
\begin{equation}
  H=\sum_j h^{\rm kin}_{j,j+1}+h^{\rm ex}_{j,j+1} .
\end{equation}
 The ``kinetic term'' $h^{\rm kin}_{j,j+1}$ generates hopping processes for particles of both colors over the vacuum state
and it is given  by
\begin{equation}
\label{xxckin}
  h^{\rm kin}_{j,j+1}=
\sum_{a=1,2}   \sigma^{+,(a)}_j\sigma^{-,(a)}_{j+1}+ 
\sigma^{-,(a)}_j\sigma^{+,(a)}_{j+1}.
\end{equation}
The ``color exchange term'' $h^{\rm ex}_{j,j+1}$ generates an exchange of colors for neighboring particles and 
is given by
\begin{equation}
  h^{\rm ex}_{j,j+1}= \gamma(
  S_j^+S_{j+1}^-+S_j^-S_{j+1}^+).
\end{equation}
For the moment, we do not specify the boundary conditions, because below we will treat both the periodic and the open boundary cases.

The Hamiltonian conserves the total particle number operators for particles of both colors, which are given by
\begin{equation}
  N^{(a)}_{\rm tot}=\sum_j N^{(a)}_j,\quad a=1,2 .
\end{equation}

The model is integrable for $\gamma=0$, in which case it becomes the Maassarani-Mathieu spin chain \cite{su3-xx}. That
model is treated in Sec. \ref{sec:MM}. 
The model is also integrable for $\gamma=1$, in which case the local Hamiltonian
density becomes the $SU(3)$-invariant permutation operator, and we obtain a spin chain solvable by the nested Bethe ansatz (see, 
e.g., \cite{Slavnov-nested-intro,fedor-lectures}). The model is generally not integrable for other values of $\gamma$. 

However, there are three invariant sectors in the Hilbert space in which it
is trivially integrable for all $\gamma$.
These three sectors can be denoted as $\HH^{(ab)}$ with $a, b=0, 1, 2$ and $a<b$, and they are
spanned by basis states that are products of the local states $\ket{a}$ or $\ket{b}$:
\begin{equation}
  \HH^{(ab)}=\bigotimes_{j=1}^L \mathrm{span}(\ket{a}_j,\ket{b}_j),\qquad 0\le a<b\le 2
\end{equation}
Alternatively, $\HH^{(01)}$ and $\HH^{(02)}$ are the sectors with $N^{(2)}_{\rm tot}=0$ and $N^{(1)}_{\rm tot}=0$, respectively, and $\HH^{(12)}$ is the sector where $N^{(1)}_{\rm tot}+N^{(2)}_{\rm tot}=L$.

It can be seen directly that for $\gamma\ne 0$, the restriction of the Hamiltonian to any of the three sub-spaces $\HH^{(ab)}$
yields the XX model, which is a spin-1/2 chain given by
\begin{equation}
  \label{HXX}
  H^{(XX)}=\sum_j  \sigma^{+}_j\sigma^{-}_{j+1}+ 
\sigma^{-}_j\sigma^{+}_{j+1}.
\end{equation}
The restriction of $H$ to the subspaces $\HH^{(01)}$ and $\HH^{(02)}$ yields the $XX$ model, because the spin exchange term $h^{\rm ex}_{j,j+1}$ vanishes identically (because only one color is present in the subspace), and the
kinetic term simplifies to \eqref{HXX}, since now only one type of $\sigma^{\pm,a}$ is present. In the case of $\HH^{(12)}$, the kinetic term of $H$ acts identically as zero
(because there is no local vacuum state in the subspace), but the color exchange terms are identical to \eqref{HXX}
multiplied by the overall factor $\gamma$. 

Quantum chaos arises in the full Hilbert space because the typical eigenstates have particles of both colors
present (all three local basis states are present), the interaction is generic, and only fine-tuned values of $\gamma$ could guarantee the solvability of the model.

The dimensions of the three subspaces grow exponentially as $2^L$, but they occupy an
exponentially small ratio of the full Hilbert space of dimension $3^L$. This means, according to Definition \ref{defweakergb}, that we obtained a model with weak
ergodicity breaking.

It is crucial that the isolated subspaces are tensor products of local
spaces, therefore we say that the embedding of the XX model is performed in a trivial manner.
Let us formalize this observation on the level of projectors. We define operators
$P^{(ab)}$ as those operators that project onto $\HH^{(ab)}$: 
\begin{equation}
  \begin{split} 
    P^{(01)}&=\prod_{j=1}^L  (1-N^{(2)}_j),\\
     P^{(02)}&=\prod_{j=1}^L  (1-N^{(1)}_j),\\
    P^{(12)}&=\prod_{j=1}^L  (N^{(1)}_j+N^{(2)}_j).\\    
 \end{split}
\end{equation}
These operators have a product form with respect to spatial bipartitions. Therefore, they have vanishing operator
entanglement for any spatial bipartition of the system.  This is the reason why we call the model a ``trivial'' example
of weak ergodicity breaking with integrable subspaces.

\subsection{More general trivial embeddings}

\label{sec:trivgen}

Trivial embeddings such as in the previous Subsection can be formalized as follows. Let $V_j=\complex^D$ denote local
Hilbert space with some $D>2$. Let us assume that there exist local subspaces $A_j\subset V_j$ with dimension
$1<D'<D$. Using these local subspaces, we can construct a distinguished sector of the Hilbert space as
\begin{equation}
\HH'=  \bigotimes_{j=1}^L A_j.
\end{equation}
Let $P'$ denote the projector to this subspace. We will also use the complement subspaces $B_j$ such that
$V_j=A_j\oplus B_j$. 
Let us then consider Hamiltonians which keep this sector invariant, which means that
\begin{equation}
  [H,P']=0.
\end{equation}
A Hamiltonian $H$ can have weak ergodicity breaking by an integrable subspace, if $H$ is chaotic, but the restriction to
the subspace given by
\begin{equation}
  P'HP'
\end{equation}
is integrable. 

Such examples can be constructed easily. One needs to choose a basis in $A_j$, and an integrable model with local
dimension $D'$. Then those matrix elements of the Hamiltonian density for $H$ which belong to the subspaces $V_j$ can be
filled up by ``copying'' the matrix elements of the corresponding integrable Hamiltonian. Next, one needs to ensure that
the full Hamiltonian has ``subspace conservation'', which is the generalization of particle number conservation. More
formally, in the case of nearest neighbor interactions, 
{the local Hamiltonian acting on sites $j$ and $j+1$} should have matrix elements only of the type
\begin{equation}
  \begin{split}
    A_j\otimes A_{j+1} \quad&\to\quad A_j\otimes A_{j+1} \\
    A_j\otimes B_{j+1} \quad&\to\quad (A_j\otimes B_{j+1})\oplus (B_j\otimes A_{j+1})\\
    B_j\otimes A_{j+1} \quad&\to\quad  (A_j\otimes B_{j+1})\oplus (B_j\otimes A_{j+1})\\
      B_j\otimes B_{j+1} \quad&\to\quad B_j\otimes B_{j+1}  \\
  \end{split}
\end{equation}
This guarantees that the integrable sector is indeed preserved. However, by choosing generic interaction terms for the
matrix of the last three types of processes above, one ends up with a generic ergodic model. The interactions in the integrable sector are given by the first type of matrix elements listed above. 
Examples in the literature which fall in this category include \cite{oitmaa2003spin,
  yang2008rigorous,sun-hubbard-eta}.

Such embeddings are trivial because the distinguished subspace is of product form. In the following sections, we treat models where the isolated integrable subspace is not of product form and the corresponding projection operators have nonzero operator entanglement.

\section{The Maassarani-Mathieu spin chain and its perturbations}

\label{sec:MM}

This is a spin chain model with local dimension $3$. 
Consider the Hamiltonian
\begin{equation}
  \label{XXCH}
    H_0 =\sum_j   h^{\rm kin}_{j,j+1}+h^{\rm diag}_{j,j+1}.
\end{equation}
Here, the kinetic term is given by \eqref{xxckin} above. 
The diagonal interaction term is
\begin{equation}
h^{\rm diag}_{j,j+1}=   
\sum_{a=1,2}
  \alpha (N^{(a)}_j P_{j+1}+P_{j}N^{(a)}_{j+1})+
  \beta N^{(a)}_j N^{(a)}_{j+1},
\label{diagtermXXX}
\end{equation}
where $\alpha$, $\beta$ are arbitrary real parameters. 
See Eqs. \eqref{eq:Naop} and \eqref{eq:Pop} for the definitions of $N^{(a)}_j$ and $P_j$.

In the case of $\alpha=\beta=0$ the model is equivalent to the Maassarani-Mathieu spin chain \cite{su3-xx}. That model is also equivalent to the so-called $t$-$0$ model, which can be obtained from the large coupling limit of the Hubbard model (see, 
e.g., \cite{caspers1989exact, ogata1990bethe, gomez1992new, murakami-goehmann-hubbard-t0,sajat-hubbard}). 
Introducing $\alpha\ne 0$ one obtains a special case of the so-called XXC models of Maassarani
\cite{XXC}, up to an irrelevant additive constant. Using the terminology of \cite{sajat-superintegrable} this is an XXC model of type $1+2$. The fully general XXC models and their nomenclature will be explained below in Sec. \ref{sec:XXC}. 
Switching on $\beta\ne 0$ we arrive at the models investigated recently in
\cite{chihiro-embedding,chihiro-naoto}.

\subsection{Irreducible strings}

We consider the model $H_0$ given by Eq. \eqref{XXCH} with an arbitrary nonzero $\alpha, \beta$. 
For simplicity, we consider the model with open boundary conditions. Following the ideas of
\cite{folded-XXZ-stochastic-1,sajat-folded,sajat-mpo}, we construct the irreducible string in the computational basis for each basis state separately. The idea is to regard the basis states as a string consisting of the numbers $0$, $1$, and $2$, and to delete every occurrence of $0$. The string obtained this way describes the relative ordering of the ``color'' of the particles, by disregarding their positions. As explained in \cite{sajat-mpo}, this string can also be seen as describing
the ``color part'' of the wave function.

The kinetic term $h_{j,j+1}^{\rm kin}$ in the Hamiltonian $H_0$ \eqref{XXCH} describes the hopping of the quasi-particles, but it is completely insensitive to the color. At the same time, two particles with different colors cannot exchange their positions. The diagonal terms given by $h^{\rm diag}_{j,j+1}$ do not affect the ordering of the particles either. This implies that the irreducible string is indeed a constant of motion. Note that we used open boundary conditions so that the irreducible string is well-defined; in the case of periodic boundary conditions, it can be defined only up to cyclic shifts.

\subsection{Integrability of the Maassarani-Mathieu chain}

\label{sec:separation}

Let us now set $\alpha=\beta=0$, therefore we consider the Hamiltonian given by the kinetic term only. This is the model treated originally in \cite{su3-xx}. The model is integrable in every fragment of the Hilbert space. 
This can be proven by embedding the Hamiltonian into the framework of the Yang-Baxter equation
\cite{su3-xx}. Alternatively, the integrability
can be demonstrated directly by a real space approach, namely by performing a color-charge separation \cite{sasha-anyon}.

Following \cite{sajat-mpo}, we parametrize the states of the Hilbert space as
\begin{equation}
  \begin{split}
  \label{statesep}
  \ket{\Psi}=&\sum_{x_1<\dots<x_N}
  \chi(x_1,x_2,\dots,x_N) \times\\ &\times \sum_{a_j=1,2}
\psi_{a_1,a_2,\dots,a_N}
\prod_{j=1}^{N}\sigma^{+,(a_j)}_{x_j}  \ket{\emptyset},
\end{split}
\end{equation}
where
\begin{equation}
  \ket{\emptyset}=\ket{0000\dots 0}
\end{equation}
is the vacuum state annihilated by $\sigma^{-,(a)}_x$ for all $x$.
In this formula, $\chi(x_1,x_2,\dots,x_N)$ is the wave function describing the charge degrees of freedom (positions of particles), while $\psi_{a_1,a_2,\dots,a_N}$ describes the color degrees of freedom.
One can always parametrize the states in this way, and generally, the color part $\psi_{a_1,a_2,\dots,a_N}$ will depend on the coordinates $x_1,\dots,x_N$, too.

It is a special property of our Hamiltonian that in the case of open boundary conditions the color part of the wave
function becomes independent of the coordinates, and it becomes a constant of motion. It can be seen that the
kinetic term of the Hamiltonian acts only on the charge part of the wave function. This happens because the hopping
terms of \eqref{xxckin} are completely insensitive to the color 
degree of freedom. If two particles occupy two neighboring positions $j$ and $j+1$, then there are no hopping events because the Hamiltonian density $h_{j,j+1}^{\rm kin}$ annihilates the state, irrespective of the color degree of freedom. 

The time evolution of the charge degree of freedom is equivalent to that of the XX model. To see this, consider an
auxiliary color-$1/2$ chain with wave functions given by
\begin{equation}
  \ket{\psi}=\sum_{x_1<\dots<x_N}
  \chi(x_1,x_2,\dots,x_N)
  \prod_{j=1}^{N}\sigma^-_{x_j}  \ket{\emptyset},
\end{equation}
where now $\ket{\emptyset}=\ket{\uparrow\uparrow\dots\uparrow}$. It can be seen directly, that if $\ket{\Psi}$ evolves
with the Hamiltonian  
$H_0$ with $\alpha=\beta=0$
and if $\chi(x_1,\dots,x_N)$ is extracted from that wave function, then the auxiliary
state $\ket{\psi}$ evolves under the action of the XX model Hamiltonian given by \eqref{HXX}.

The color part of the wave function $\ket{\Psi}$ can be identified with the
irreducible string defined above. To be more precise, if the wave function $\psi_{a_1,a_2,\dots,a_N}$ is equal to 1 for a specific sequence $a_1,a_2,\dots,a_N$, and zero otherwise, then the given sequence is identical to the irreducible string.  
We should note that the conservation of the irreducible string and the exact color-charge separation led to the exact computation of anomalous fluctuations in closely related cellular automata (see, 
e.g., \cite{prosen-MM-anomalous,prosen-anomalous-universal}).
The exact color-charge separation also leads to the existence of a large family of MPO symmetries; for details see Ref. \cite{sajat-mpo}.

\subsection{Non-integrable perturbations}

\label{sec:nonintxxc}

Now we switch on the coupling constants $\alpha, \beta$ in $h^{\rm diag}_{j,j+1}$. This is the model treated in Ref. \cite{chihiro-embedding}. 
Direct investigation of the two terms in $h^{\rm diag}_{j,j+1}$ shows that the color-charge separation is kept intact even in the case of a nonzero $\alpha$, and only the addition of $\beta\ne 0$ breaks this property. This happens because the $\alpha$-dependent diagonal term is again insensitive to the color degree of freedom. On the other hand, the $\beta$-dependent term does make a distinction between colors, because it gives a nonzero contribution only when two neighboring excitations have the same color. Therefore, the $\beta$-dependent term breaks the color-charge separation. 

It follows that the Hilbert space fragmentation is kept intact also for $\beta\ne 0$, but integrability is broken in almost all fragments. There will be only a few
number of selected
fragments (in the case of open boundary conditions), in which the model remains integrable. Interestingly, we find that
some of these
subspaces do not take a product form, thus the embedding of the integrable subspace is performed on an entangled basis.

Now we treat the individual integrable subspaces separately.

{\bf Subspace $\HH^{(12)}$}. In this subspace, the model is trivially integrable: the kinetic part acts as identically
zero, and the diagonal part is integrable, but it does not generate any dynamics.

{\bf Subspaces $\HH^{(01)}$ and $\HH^{(02)}$}. In these two subspaces, there is only one excitation present, and the irreducible string consists only of $1$'s and $2$'s, respectively. We show that, in these two subspaces, the model is equivalent to the XXZ model with some anisotropy parameter.

We consider, for example, $\HH^{(01)}$, and identify this subspace as the standard color-$1/2$ Hilbert space, with the identification of basis states as $\ket{0}=\ket{\uparrow}$ and $\ket{1}=\ket{\downarrow}$. The Hamiltonian is then
written as
\begin{equation}
  \begin{split}
  H_0=&\sum_j
   \sigma^{+}_j\sigma^{-}_{j+1}+ 
   \sigma^{-}_j\sigma^{+}_{j+1}+\\
   &+\alpha (N_jP_{j+1}+P_{j}N_{j+1}) +\beta  N_j N_{j+1},
  \end{split}   
\end{equation}
where now
\begin{equation}
  N=\ket{1}\bra{1},\qquad P=\ket{0}\bra{0}.
\end{equation}
Using the identities
\begin{equation}
  N+P=1,\quad N-P=Z,
\end{equation}
together with the expression of the raising/lowering operators with the local Pauli operators $\{ X_j, Y_j, Z_j \}$,
we can rewrite the Hamiltonian as
\begin{equation}
  \begin{split}
  H_0=&\sum_j
\frac{1}{2}  ( X_jX_{j+1}+ 
   Y_jY_{j+1})+\\
   &
+(-2\alpha+\beta)Z_jZ_{j+1}+2\beta Z_j+2\alpha+\beta.
 \end{split}   
\end{equation}
This way, we obtained the XXZ model with anisotropy $\Delta=2(-2\alpha+\beta)$ and a longitudinal magnetic field
$h=4\beta$.
The restriction of the full Hamiltonian to the subspace $\HH^{(02)}$ yields the same specialization of the XXZ model.

{\bf Entangled subspaces.} Interestingly, this model has additional integrable subspaces, which are embedded in an
entangled basis. These are the fragments that have a strictly alternating irreducible string. This means that
neighboring excitations always have different colors. In the case of open boundary conditions, we can have an arbitrary number of excitations (up to the length of the chain), but in the case of periodic boundary conditions, only an even number of excitations are allowed. In the periodic case, the irreducible string becomes $121212\dots 12$, with an
arbitrary number of repetitions of $12$. We will denote the subspace with $n=2k$ particles with alternating colors as
$\tilde\HH_n$, and we also define the direct sum of these subspaces as
\begin{equation}
  \tilde \HH=\mathop{\mathop{\bigoplus}_{n=2k}}_{n<L} \tilde \HH_n
\end{equation}

Let us now compute the restriction of the Hamiltonian to these subspaces. For simplicity, we consider only the periodic
case. We identify the local vacuum state with $\ket{\uparrow}$ and the excitations with $\ket{\downarrow}$.
Inspection shows that within this subspace, the Hamiltonian density is insensitive to the color of the excitations,
thus we obtain a color-charge separation in this sector once again.
We find that the kinetic part acts in the same way as before, but
the action of the diagonal part will be different from the previous case. We get
\begin{equation}
  \begin{split}
  H_0=&\sum_j
   \sigma^{+}_j\sigma^{-}_{j+1}+ 
   \sigma^{-}_j\sigma^{+}_{j+1} 
   \\
   &+\alpha (N_jP_{j+1}+P_{j}N_{j+1}),
  \end{split}   
\end{equation}
This happens because the $\beta$-dependent term in the original Hamiltonian is sensitive only to those configurations
where two excitations of the same color take two neigboring positions, and such local configurations do not occur in this subspace. We also used the fact that neither the kinetic term nor the $\alpha$-dependent diagonal term makes any distinction between excitations with different colors. We can rewrite this Hamiltonian as
\begin{equation}
  \begin{split}
  H_0=&\sum_j
\frac{1}{2}  ( X_jX_{j+1}+ 
   Y_jY_{j+1})+\\
   &-2\alpha Z_jZ_{j+1}+2\beta.
 \end{split}   
\end{equation}
This way, we obtained the XXZ model with anisotropy $\Delta=-4\alpha$ and zero longitudinal magnetic field.

The defining property of the subspace $\tilde \HH$ is that consecutive particles must have different colors, and this is a
nonlocal constraint. Therefore, the embedding of the integrable XXZ model is now performed on an entangled basis.
We can also construct the projector onto the full subspace $\tilde \HH$ in the form of an MPO. More precisely, we obtain an MPO that is almost identical to the projector, 
except for an additional term, which is needed for technical reasons.

First, we define a
periodic MPO as 
\begin{equation}
  \label{MPO}
\tilde P=  \text{Tr}_a [\La_{L,a}\dots\La_{2,a}\La_{1,a}],
\end{equation}
where $\La_{j,a}$ is a linear operator acting on the tensor product of the local space at site $j$ and an auxiliary
vector space indexed by $a$. In our case, this auxiliary space is a spin-$1/2$ space. We find that the desired projector is
obtained by
\begin{equation}
  \La_{j,a}=N^{(1)}_j\sigma^+_a+N^{(2)}_j\sigma^-_a+P_j.
\end{equation}
Multiplying these operators, expanding the product and taking the trace in \eqref{MPO}, we find that only those
combinations survive that have an alternating sequence of $N^{(1)}_j$ and $N^{(2)}_j$, with an arbitrary number of insertions of the operators $P_j$. Indeed, this implies that the irreducible string becomes $121212\dots 12$. An exception is given by the reference state, in which case neither $N^{(1)}_j$ nor $N^{(2)}_j$, appears, and due to the trace in the auxiliary space, we obtain an overall factor of $2$. 
This implies that the desired projector is given by
\begin{equation}
\label{kivonas}
    \tilde P - \ket{\emptyset}\bra{\emptyset},
\end{equation}
where $\ket{\emptyset}=|00\dots 0 \rangle$. One could rewrite this operator as an MPO with enlarged bond dimension, but we disregard this irrelevant technical detail.

Despite the fact that the model is chaotic in almost all subspaces, one can argue that it still has strong ergodicity
breaking. The Hilbert space fragmentation is kept intact by the non-integrable perturbation, and the model has strong
fragmentation: the dimensions of the fragments do not grow faster than $2^L$ while the full Hilbert space has dimension
$3^L$. 

\subsection{Weak ergodicity breaking with integrable subspaces}

Now we show a new mechanism that leads to weak ergodicity breaking. The idea is to add a perturbation which will undo the fragmentation by kinetically connecting the various sectors, 
while leaving the integrable sector $\tilde \HH$ intact. 
To this end, we take the Hamiltonian
\begin{equation}
  \label{XXCH2}
    H=\sum_j   h^{\rm kin}_{j,j+1}+h^{\rm diag}_{j,j+1}+h^{\rm pair}_{j,j+1}
\end{equation}
with the perturbation
\begin{equation}
h^{\rm pair}_{j,j+1} = \gamma
S_j^-S_{j+1}^-+ \gamma^* S_j^+S_{j+1}^+, 
\label{gammatermXXX}
\end{equation}
where the parameter $\gamma$ is an arbitrary complex number. This term can be interpreted as a ``color-flip'' operation performed simultaneously on two excitations if they occupy adjacent sites. 
Physically, this term can be viewed as a spin-spin interaction in the corresponding $t$-$J$-like model. When $\gamma$ is a real parameter, $h^{\rm pair}_{j,j+1}$ describes the anisotropy in the XY plane. On the other hand, when $\gamma$ is purely imaginary, this term takes the form of the symmetric off-diagonal exchange interaction, 
often referred to as the $\Gamma$-term in the context of Kitaev materials \cite{rau2014generic}.

The perturbation preserves the total particle number
\begin{equation}
  N^{\rm tot}=N^{(1)}+N^{(2)},
\end{equation}
but the individual particle numbers for the two colors will not be preserved anymore. It can be seen directly that the perturbation leaves the distinguished subspace $\tilde \HH$ invariant; in fact, the
action of $h^{\rm pair}_{j,j+1}$ vanishes on this subspace. This follows from the fact that the spin-flip operations in $h^{\rm pair}_{j,j+1}$ require to have two excitations of the same color on neighboring sites, and such configurations
are forbidden in $\tilde \HH$. As an effect, the restriction of the full Hamiltonian to the subspace $\tilde \HH$ is
independent of $\gamma$, therefore it is still described by the integrable XXZ model. 
At the same time, the perturbation destroys fragmentation for all the other subspaces.

\section{The XXC models and their perturbations}

\label{sec:XXC}

Here we review the XXC models \cite{XXC}, which are generalizations of the Maassarani-Mathieu chain treated in the previous Section.  They are all integrable, they all have Hilbert space fragmentation, and their perturbations can be tailored to have weak ergodicity breaking with isolated integrable sectors. It will be shown later in Sec. \ref{sec:ladder} that the so-called XX spin ladder treated earlier in \cite{znidaric-coexistence} is a perturbation of an XXC model.

Let $D\ge 2$ be the dimension of the local spaces in our model, and let us choose two integers $1<d_1\le d_2$ such that $D=d_1+d_2$. We also define the sets $A=\{1,2,\dots,d_1\}$, $B=\{d_1+1,d_1+2,\dots,D\}$.
We define a kinetic term and a diagonal interaction term via the Hamiltonian densities $h^{\rm kin}_{j}$ and $h^{\rm diag}_{j}$:
\begin{equation}
    h^{\rm kin}_j=\sum_{a\in A}\sum_{b\in B} 
    \ket{ab}\bra{ba}+ {\rm H.c.} .
\end{equation}
Here the dependence on the site index $j$ is suppressed on the rhs. 
The diagonal term is
\begin{equation}
    h^{\rm diag}_j=(h^{\rm kin}_j)^2=
    \sum_{a\in A}\sum_{b\in B}
    (\ket{ab}\bra{ab}+\ket{ba}\bra{ba}).
\end{equation}
Finally, the full Hamiltonian is
\begin{equation}
\label{XXCH0}
    H^{\rm int}=\sum_j h^{\rm kin}_j-\Delta h^{\rm diag}_j
\end{equation}
where $\Delta\in\valos$ is a parameter analogous to the anisotropy of the XXZ spin chain.

Following \cite{sajat-superintegrable}, we call the model defined by \eqref{XXCH0} the XXC model of type $d_1+d_2$ with anisotropy $\Delta$. 
In the case of $d_1=d_2=1$ one obtains precisely the spin-$1/2$ XXZ chain with anisotropy $\Delta$.
The Maassarani-Mathieu (MM) chain treated in Sec. \ref{sec:MM} can be seen as the XXC model of type $1+2$ with $\Delta=0$.
The XXC models of type $1+d$ with some $d>2$ are simple generalizations of the MM chain. 

For further generalizations of the XXC models, see Ref. \cite{multiplicity-models} and the recent work \cite{marius-flag}. The models treated in \cite{multiplicity-models} could be used to embed integrable spin chains with higher rank symmetries into otherwise chaotic models. Exploring all those possibilities is beyond the scope of the present work.

\subsection{Hilbert space fragmentation}

The XXC models have a fragmented Hilbert space. 
The models of type $1+d$ possess one irreducible string, which is obtained in the computational basis by deleting every occurrence of $1$ from the sequence of local states. 
In contrast, the general XXC models with $d_2\ge d_1> 1$ possess two irreducible strings. The two irreducible strings are obtained in the computational basis by deleting every occurrence of local states from the set $A$ and $B$, respectively. For example, let us take $d_1=d_2=2$ and consider the basis state
\begin{equation}
\ket{12134133241}
\end{equation}
The two irreducible strings are
\begin{equation}
 34334 \quad\text{and}\quad   121121
\end{equation}
Conservation of {\it both strings} follows from the fact that  
the $h^{\rm kin}_j$ do not swap two local states that are both from $A$ or both from $B$.
Thus, the relative order of the local states belonging to both sets is conserved.

\subsection{Integrability and color-charge separation}

The integrability of the general XXC models was proven in \cite{XXC}. The proof can be given via algebraic means, but perhaps a more transparent proof is to consider the dynamics of the model. Similar to the treatment in Sec. \ref{sec:separation}, it is possible to perform a color-charge separation. However, in this case, we have two ``color'' degrees of freedom corresponding to the subsets $A$ and $B$ of local states. In the computational basis, the charge part of the wave function simply just describes whether we place a local state from either the $A$ or the $B$ set, and then the color parts describe the internal degrees of freedom for both sets.

Now we formalize this separation of variables. Consider the expansion of the wave function into the computational basis
\begin{equation}
    \ket{\Psi}=\sum_{k_1,k_2,\dots,k_L=1}^D \Psi_{k_1,k_2,\dots,k_L}
    \ket{k_1,k_2,\dots,k_L},
\end{equation}
where $L$ denotes the length of the spin chain.
For each basis state, we compute three sequences of numbers. Two sequences will be given by the two irreducible strings; let us denote them by $I_A$ and $I_B$. The third sequence will correspond to the ``charge part of the wave function'', and will be denoted as $K$. The procedure to obtain these sequences is the following. We proceed along the sites from $j=1$ to $j=L$, starting with empty sequences. For every $j$, we consider the local state $k_j$. If $k_j\in A$, then we append a $0$ to the sequence $K$ and append $k_j$ to the sequence $I_A$. If $k_j\in B$, then we append a $1$ to the sequence $K$ and append $k_j-d_1$ to the sequence $I_B$. Note that the resulting sequence $K$ has length $L$, but the lengths of $I_A$ and $I_B$ depend on the specific states. Nevertheless, we always have $L=|I_A|+|I_B|$.

The color-charge separation means that the wave function is factorized as
\begin{equation}
  \Psi_{k_1,k_2,\dots,k_L}=\chi^{(A)}(I_A) \chi^{(B)}(I_B) \chi^{\rm ch}(K)
\end{equation}
Here $\chi^{(A)}(I_A)$ and $\chi^{(B)}(I_B)$ are two wave functions that depend on the irreducible strings $I_A$ and $I_B$, and $\chi^{\rm ch}(K)$ is the wave function for the charge part. 
If we consider the model defined by the kinetic terms alone, then we find that the charge part of the wave function evolves according to the XX model in Eq. \eqref{HXX}, while the two color parts are constants of motion. If we also add the diagonal interaction terms, then we find that the charge part evolves according to the XXZ model with anisotropy $\Delta$, while the color parts are still constants of motion.

\subsection{Perturbations with weak ergodicity breaking}

It is possible to construct perturbations of the XXC models, which would destroy the color-charge separation and undo the fragmentation for almost all of the original sectors, meanwhile keeping one or a few selected sectors untouched and therefore integrable. The idea is to select a desired pattern in the computational basis, and to choose new interaction terms which act identically as zero on the specifically chosen pattern.

This approach might seem somewhat unnatural, given that we are considering higher-dimensional local spaces, and the resulting models might appear even more exotic than our other examples. However, perhaps surprisingly, we find that the so-called XX spin ladder (treated in the next section) falls precisely in this category. That spin ladder model is rather natural and yet it can be described by the abstract framework that we laid out in this section.

\section{The XX spin ladder}

\label{sec:ladder}

Here we treat the so-called XX spin ladder model investigated in \cite{znidaric-coexistence}. This is a spin ladder model, and to write down the Hamiltonian we will use the local operators $\sigma^{x,y,z}_j$ and $\tau^{x,y,z}_j$, which correspond to Pauli matrices $X, Y, Z$ acting on the two legs of the ladder. 
The Hamiltonian is 
\begin{equation}
H=\sum_j  h^{||}_j+\sum_j h_j^{\perp}
\end{equation}
with
\begin{equation}
    h^{||}_j=
    \sigma^x_{j}\sigma^x_{j+1}+
    \sigma^y_{j}\sigma^y_{j+1}+
    \tau^x_{j}\tau^x_{j+1}+
    \tau^y_{j}\tau^y_{j+1}
\end{equation}
and
\begin{equation}
    h_j^{\perp}=
    J( \sigma^x_{j}\tau^x_{j}+
    \sigma^y_{j}\tau^y_{j}+
 \tilde \Delta   \sigma^z_{j}\tau^z_{j}) .
\end{equation}
In a special limit, this model reproduces the Hamiltonian of the Hubbard model \cite{shastry-r}. If we identify the local spaces of the two legs of the ladder with the fermionic states of two different spins (via the Jordan-Wigner transformation), then the special limit 
$J\to 0$, $\tilde\Delta\to\infty$ with $J \tilde\Delta\to U$ reproduces the Hubbard model with on-site interaction parameter $U$. However, for generic values of $J$ and $\tilde \Delta$ the model is different, and it is not integrable.

Nevertheless, it was shown in \cite{znidaric-coexistence} that this model also has an integrable sub-space. In this work, we show that the mechanism of weak ergodicity breaking in this model
is very similar to the one discussed in our previous examples. Namely, the model Hamiltonian can be seen as a
perturbation of a model with Hilbert space fragmentation, such that the perturbation connects many sectors but it keeps
a selected integrable sector invariant. 

\subsection{Connection with the XXC models}

Following \cite{znidaric-coexistence}, we investigate the various matrix elements of the Hamiltonian density. This will lead to a rewriting which will make the perturbed fragmentation apparent. 

The rung configurations can be described according to their $SU(2)$ eigenvalues. There are two states with zero total
magnetization on the rung, these are the singlet and triplet states
\begin{equation}
  \begin{split}
  \ket{S}&=\frac{1}{\sqrt{2}}(\ket{\uparrow\downarrow}-\ket{\downarrow\uparrow})\\
 \ket{T}&=\frac{1}{\sqrt{2}}(\ket{\uparrow\downarrow}+\ket{\downarrow\uparrow})
\end{split}
\end{equation}
Furthermore, there are the additional two states from the triplet, which we denote as
\begin{equation}
  \ket{-}=\ket{\downarrow\downarrow},\qquad \ket{+}=\ket{\uparrow\uparrow}.
\end{equation}
These local rung states are eigenstates of the corresponding rung Hamiltonian density $h^\perp_j$. 

It was shown in \cite{znidaric-coexistence}  that the Hamiltonian densities $h^{||}_j$ can be represented by the following matrix
elements in the rung eigenbasis:
\begin{equation}
  \begin{split}
    h^{||}_j=&2\sum_{a=+,-}\sum_{b=S,T} \ket{ab}\bra{ba}+ {\rm H.c.}\\
  &+4 \big(\ket{TT}-\ket{SS}\big)\big(\bra{+-}+\bra{-+}\big)+ {\rm H.c.},
  \end{split}
\end{equation}
where the dependence on the site index $j$ will be suppressed on the rhs.
Now we introduce two separate definitions for the two types of matrix elements above, namely,
\begin{equation}
  h^{||}_j=h^{\rm kin}_j+h^{X}_j,
\end{equation}
where
\begin{equation}
  \begin{split}
    h^{\rm kin}_j&=  2\sum_{a=+,-}\sum_{b=S,T} \ket{ab}\bra{ba}+ {\rm H.c.},\\
    h^{X}_j&=4 \big(\ket{TT}-\ket{SS}\big)\big(\bra{+-}+\bra{-+}\big)+{\rm H.c.}.
    \end{split}
\end{equation}
The kinetic terms in $h^{\rm kin}_j$ describe special types of hopping processes: the configurations $+$ and $-$ can exchange
places with the configurations $S$ and $T$, but a $+$ and a $-$ cannot cross each other, and similarly for $S$ and $T$. This means that even though the positions of the configurations do change, the sequences of the $+$ and $-$ states, and also of the $S$ and $T$ states are preserved by time evolution. This means that $h^{\rm kin}_j$ preserves two
separate irreducible strings. In fact, the model defined by
\begin{equation}
  H^{\rm kin}=\sum_j h^{\rm kin}_j
\end{equation}
is one special example of the more general XXC models defined in \cite{XXC}. Using the nomenclature 
introduced in the previous section,
this is the XXC model of type $2+2$, with interaction parameter $\Delta=0$. The two irreducible strings for this model are the two sequences formed from the letters $+$ and $-$, and also $T$ and $S$.

Altogether we found that the XX spin ladder model can be seen as the XXC model of type $2+2$ perturbed by ``magnetic fields'' given by
$h^\perp_j$ and also by the additional operator
\begin{equation}
  H^X=\sum_j h^X_j.
\end{equation}

\subsection{Integrable subspace of the perturbed model}

It was explained in \cite{znidaric-coexistence} that the XX model can be embedded into the perturbed model. There are two possible embeddings. The irreducible string corresponding to the letters $S$ and $T$ 
is 
\begin{equation}
    STSTST \cdots 
\end{equation}
in both cases. Regarding the letters $+$ and $-$, one needs to choose a ferromagnetic configuration, i.e. the corresponding irreducible string has to be $++++\cdots$ or $----\cdots$.

This is a good choice, because a $+$ or a $-$ can propagate in the background formed by any sequence of the $S$ and $T$ letters, but if the irreducible strings are chosen as described above, then the additional terms $h^X_j$ will act as identically zero. This happens because they act nontrivially only on local two-site configurations $+-$, $-+$, $TT$, and $SS$, and these do not occur if we choose the irreducible strings as described above.
 
\section{The folded XXZ model and its perturbations}

In this section, we consider the folded XXZ model and its variations. We will obtain a different mechanism for embedding
an integrable sector into a chaotic model.

The folded XXZ model is an integrable spin-1/2 chain, which was studied in \cite{folded1,folded2,sajat-folded}, although
the model and certain perturbations appeared earlier in \cite{folded0a,folded0b,folded-XXZ-stochastic-1}.

The integrable Hamiltonian is given by
\begin{equation}
  \label{foldede}
  H_{\rm folded}=\sum_j \frac{1+Z_jZ_{j+3}}{2} (X_{j+1}X_{j+2}+Y_{j+1}Y_{j+2}).
\end{equation}
As discussed in \cite{folded1,folded2,sajat-folded}, one can perform a so-called bond-site transformation to arrive at
a dual Hamiltonian
\begin{equation}
  \label{foldedb0}
    H^{\rm int}=\sum_{j} h^{\rm kin}(j)
\end{equation}
where now
\begin{equation}
  \label{foldedb}
  h^{\rm kin}(j)=  (X_jX_{j+2}+Y_jY_{j+2})N_{j+1}.
\end{equation}

The duality transformation between the Hamiltonians given by \eqref{foldede} and \eqref{foldedb0}-\eqref{foldedb} was
discussed in detail 
in \cite{folded1,folded2,sajat-folded}  and it is not relevant for our purposes. We will simply regard the Hamiltonian defined by \eqref{foldedb0}-\eqref{foldedb} as our model Hamiltonian.

The model \eqref{foldedb0}-\eqref{foldedb} is a special case of the integrable spin chain with two- and three-spin interactions studied by Bariev \cite{bariev-model}. Previous work has shown that the model exhibits Hilbert space fragmentation \cite{sajat-folded}.
Below, we will introduce integrability breaking terms which also undo the fragmentation. However, the additional terms will be engineered so that
they keep a selected sector intact, not affecting integrability in that sector.

The model defined above is also related to
the Maassarani-Mathieu spin chain defined in the previous section. There is a highly nonlocal duality transformation
that connects these two models \cite{sajat-folded}. This duality will not be used below, therefore we refrain from giving the details.

\subsection{Integrability and Hilbert space fragmentation}

Here we describe the mechanism of Hilbert space fragmentation in this model. We will also give some details about the
solution of the model in a selected fragment. The model is integrable in every fragment, and for a complete treatment of
the eigenstates we refer to \cite{sajat-folded}.

The kinetic term \eqref{foldedb} generates the transitions
\begin{equation}
  \ket{\downarrow\downarrow\uparrow} \quad\leftrightarrow \quad  \ket{\uparrow\downarrow\downarrow}.
\end{equation}
One possible interpretation is that we regard two neighboring down spins as an excitation, and then the kinetic term
generates a translation (or hopping) of this excitation by one site. It was argued in \cite{sajat-folded,sajat-hardrod}
that this excitation is analogous to the hard rods known from classical mechanics \cite{hard-rod-gas}, and that folded
XXZ model should be regarded as some sort of ``hard rod deformation'' of the XX model.

In contrast to the hard rods of length two, an isolated down spin embedded into a sea of up spins is immobile on its
own. However, such isolated down spins can be displaced via a scattering on the hard rods. This is described in detail
in \cite{sajat-folded}.

The fact that isolated down spins are not mobile leads to Hilbert space fragmentation. The different sectors can be
labeled by the positions of the down spins, taking into account the displacements caused by the possible scattering
processes. 

\subsection{The sector of the constrained XXZ model}

\label{sec:foldedint}

Now we focus on a special sector of the model.
This sector consists of those states in the computational basis, where every block of consecutive down spins (preceded
and followed by an up spin) has even length. If we take the pairs of the neighboring down spins and we regard the
position of (say) the first down spin as a coordinate of a particle, then we get the condition that the particle
coordinates have to differ by at least $2$. This corresponds to the so-called Rydberg constraint (see Subsection \ref{sec:constr} below). Furthermore, it can be
shown that in this sector the Hamiltonian acts in the same way as the so-called constrained XXZ model (with
specification $\Delta=0$, see \cite{constrained1,constrained2,constrained3}).

This sector does not have a product space structure. We show that 
the projector to this sector is given again by an MPO.
Consider for simplicity the case with periodic boundary conditions when the length $L$ of the chain is odd.
In this case, we can construct an MPO with bond
dimension $2$. This MPO can be written in the form \eqref{MPO}, where now
\begin{equation}
  \La_{j,a}=N_jX_a+P_j P_a.
\end{equation}

This MPO will ensure that every block of consecutive down spins has an even length. Every such block has up spins
at their boundaries where the local projector $P_a$ will enforce that the bond spin is also in the up position. If there are an odd number of down spins in the block, then we would get an odd power of the spin-flip operator $X_a$ acting in
the auxiliary space, eventually giving zero coefficient for the given configuration. However, if all such blocks are
of even length, then every such state gets a coefficient equal to one. 
This argument does not work if
the length of the spin chain is even, because then the state with all spins down obtains a coefficient of $2$, which
means that the resulting MPO is not a projector; this is similar to the situation in the Maassarani-Mathieu chain, see Eq. \eqref{kivonas}. Similar to the case of the operator in \eqref{kivonas}, we could rewrite the resulting projector as an MPO with enlarged bond dimension, but we choose to disregard this subtlety, because it is not relevant to our
discussion.

Let us now discuss the Bethe ansatz solution of the model in this specific sector, following the treatment of \cite{sajat-folded}.
We introduce the local creation operators
\begin{equation}
  \label{Aa}
  A_j=
    \sigma^-_j\sigma^-_{j+1},
\end{equation}
which create two neighbouring down spins on the reference state with all spins up.
We introduce a set of momenta $\{p_1,p_2,\dots,p_N\}$, and we use the reference state $\ket{\emptyset}$ which is the ferromagnetic state with all spins up. The wave function is then written as
\begin{equation}
  \label{bondpsi}
  \begin{split}
      \ket{\Psi}=
  \sum_{x_1\le x_2\le \dots \le x_{N}}
 \sum_{\mathcal{P}\in \mathfrak{S}_{N}} e^{i \sum_{j=1}^{N} p_{\mathcal{P}_j} x_j}  \\
\times \mathop{\prod_{j<k}}_{\mathcal{P}_j>\mathcal{P}_k}
S(p_{j},p_{k})
\prod_{j=1}^{N} A_{x_j} \ket{\emptyset}.
 \end{split}
\end{equation}
Here $\mathfrak{S}_N$ stands for the permutation group of $N$ elements.
The scattering factor is given by
\begin{equation}
    S(p_1,p_2)=-e^{-i(p_1-p_2)}.
\end{equation}
The Bethe equations for a state with $N$ particles in a volume of $L$ and periodic boundary conditions are
\begin{equation}
    e^{ip_jL}\prod_{k\ne j}S(p_j,p_k)=1,\qquad j=1,\dots,N.
\end{equation}
After substitution, we get rather simple equations. Introducing the total momentum as
\begin{equation}
\label{Psum}
P^{({\rm tot})}=\sum_{j=1}^N p_j
\end{equation}
we get
\begin{equation}
    e^{iP^{({\rm tot})}L}=1,\qquad  e^{ip_j(L-N)}=(-1)^{N-1}e^{-iP^{({\rm tot})}}.
\end{equation}
An explicit solution to these equations can be given as follows. Let $\{I_j\}_{j=1,\dots,N}$ denote a set of integers, and let
\begin{equation}
    I=\sum_{j=1}^N I_j.
\end{equation}
Then we have
\begin{equation}
    P^{({\rm tot})}=\frac{\pi (2I+N(N-1))}{L}
\end{equation}
and
\begin{equation}
    p_j=\frac{(N-1+2I_j)\pi-P^{({\rm tot})}}{L-N}.
\end{equation}
The energy of the corresponding state becomes
\begin{equation}
    E=\sum_{j=1}^N \eps(p_j),\qquad \eps(p)=4\cos(p).
    \label{solutionfromBA}
\end{equation}
A more general Bethe ansatz for states with isolated down spins was also given in \cite{sajat-folded}, but we will not use the general formulas.

\subsection{Weak ergodicity breaking}

\label{sec:foldedpert}

Now we add perturbation terms to the integrable Hamiltonian $H^{\rm int}$ in Eq. \eqref{foldedb0}, to undo the fragmentation, yet keeping the selected integrable subspace intact. The key idea is to generate dynamics for the isolated down spins. Furthermore, we also break
particle conservation. However, the perturbing terms have built in controls which will annihilate them once they act on the integrable sector. 

The Hamiltonian we consider is
\begin{equation}
  \label{foldedpert}
    H=\sum_{j} h^{\rm kin}(j)+\tilde h^{\rm kin}(j)+h^X(j),
\end{equation}
where $h^{\rm kin}(j)$ is given by \eqref{foldedb}, and 
the two perturbations are
\begin{equation}
  \label{foldedx1}
   \tilde h^{\rm kin}(j)=\kappa P_j (X_{j+1}X_{j+2}+Y_{j+1}Y_{j+2})P_{j+3}
\end{equation}
and
\begin{equation}
  \label{foldedx2}
    h^X(j)=\kappa'\left[ P_jN_{j+1}P_{j+2}X_{j+3}+X_j P_{j+1}N_{j+2}P_{j+3}\right],
\end{equation}
with $\kappa, $ and $\kappa'$ real parameters.

In the sector with no isolated down spins, these terms act as identically zero. However, $\tilde h^{\rm kin}(j)$ undoes most of the fragmentation while keeping particle number conservation, and finally $h^X(j)$ also breaks particle number conservation. 

It is important that adding only $\tilde h^{\rm kin}(j)$ or $h^X(j)$ does not make the model fully ergodic, because certain nonlocal conservation laws would still remain. It follows directly from the form of these terms that $\tilde h^{\rm kin}(j)$ preserves the number of blocks of down spins with odd length, but it allows for a split or combination of a block of even length and an isolated down spin. In contrast, $h^X(j)$ preserves the number of blocks of consecutive down spins. Adding both terms will completely destroy these nonlocal conservation laws.

\subsection{Numerical investigations}

\label{sec:foldednum}

\begin{figure}
    \centering
\includegraphics[width=0.8\linewidth]{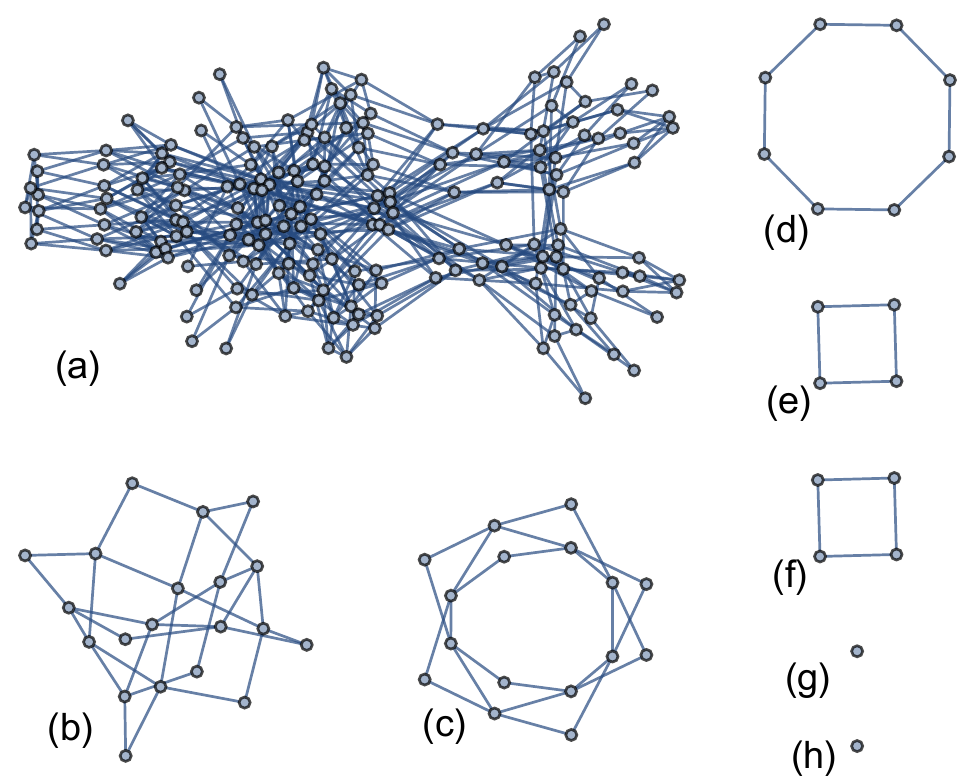}
\caption{Adjacency graph for the Hamiltonian \eqref{foldedpert} with $\kappa=\kappa'=0.789$ and $L=8$. {(a) chaotic sectors with $202$ states; (b) integrable sector with $20$ states with 2 pairs of nearby spin down particles; (c) integrable sector with $16$ states and 3 pairs of nearby spin down particles; (d) integrable sector with $8$ states, one pair of spin down; (e-f) trivially solvable sectors, $4$ states, one spin up; (g-h) $1$ state, all spins up or all spins down.}}
\label{adjacencygraphfoldedxxz}
\end{figure}
\begin{figure}
\vspace{0.5cm}
    \centering
\includegraphics[width=0.9\linewidth]{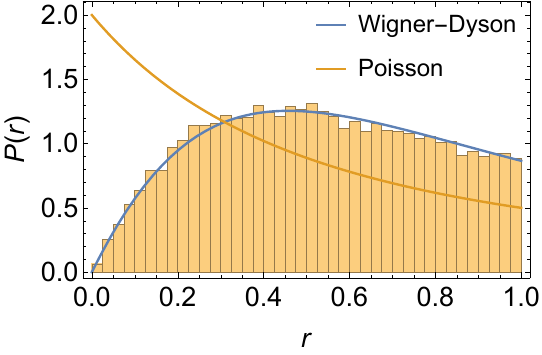}
    \caption{Distribution of the adjacent-gap ratios of the Hamiltonian in Eq. \eqref{foldedpert} for $L=20$ with $\kappa=\kappa'=0.789$. The data are collected from the subspace of the chaotic sector, which is 
    invariant  
    under translation and spatial inversion. The dimension of this subspace is $26556$. 
    The GOE Wigner-Dyson and Poisson distributions are shown for comparison.} 
\label{folded_XXZ_L20_level}
\end{figure}

In this subsection, we numerically demonstrate our main claim, namely that the Hamiltonian $H$ in Eq. \eqref{foldedpert} describes a model with weak ergodicity breaking. It is clear from the analytic derivations that the model has ergodicity breaking due to the embedded integrable subsector. Therefore, the numerical tests serve to show that this ergodicity breaking is indeed weak. In other words, we demonstrate that the majority of the eigenstates are still thermal, and they satisfy the ETH. To this end, we use the software {\it Mathematica} to perform four different numerical tests.

First, we compute the Hamiltonian's adjacency graph for the states in the computational basis. {In this graph, each vertex corresponds to a computational basis state (a product state in the $Z$ basis), and an edge represents a pair of basis states for which the Hamiltonian (with nonzero $\kappa$ and $\kappa'$) has a nonzero matrix element.} 
The results for length $L=8$ are shown in Fig. \ref{adjacencygraphfoldedxxz}. This figure confirms the presence of a large sector and some small sectors. We identify the largest connected component with the chaotic sector. The small connected components can be identified with subspaces of the integrable sector, corresponding to the different particle numbers, which are conserved in that specific sector. We recall that, in this context, "one particle" corresponds to two consecutive down spins. Two of the small connected components (e-f) in Fig. \ref{adjacencygraphfoldedxxz} do not belong to any of the integrable subspaces that we identified. However, these subsectors are trivially solvable since they correspond to states with one 
up spin on the even or the odd sites of the chain.

As a second numerical check, we computed the adjacent-gap ratio distribution, which is a standard diagnostic of ergodicity \cite{oganesyan2007localization, atas2013distribution}. {Let $E_n$ ($n=1,...,D_{\rm sub}$) be the energy eigenvalues of the Hamiltonian in ascending order. Here, $D_{\rm sub}$ is the dimension of the subspace of interest. We define adjacent gaps as $\delta_n = E_{n+1} - E_n$ ($n=1,...,D_{\rm sub}-1$). Then the adjacent-gap ratio is defined as
\begin{align}
    r_n = \frac{{\rm min} (\delta_{n+1},\delta_n)}{{\rm max} (\delta_{n+1}, \delta_n)},
\end{align}
where $n=1,...,D_{\rm sub}-2$. For non-integrable models described by GOE random matrices, the probability distribution of $r_n$ is given by $P_{\rm GOE}(r)=\frac{27}{4}\frac{r+r^2}{(1+r+r^2)^{5/2}}$, with a mean value of $\langle r \rangle_{\rm GOE} =4-2 \sqrt{3} \simeq 0.536$. On the other hand, for integrable models, the probability distribution follows $P_{\rm Poisson} (r) = \frac{2}{(1+r)^2}$, with a mean gap ratio of $\langle r \rangle_{\rm Poisson}=2\ln 2 - 1 \simeq 0.386$.}

To evaluate $r_n$ for the Hamiltonian \eqref{foldedpert}, we pick the sector of the largest dimension, which we assume to be chaotic. We further restrict it to the subspace spanned by states invariant under translation and spatial inversion, thereby removing degeneracies due to spatial symmetries. Figure \ref{folded_XXZ_L20_level} shows the distribution of $r_n$ for $L=20$ with $\kappa = \kappa' = 0.789$. Clearly, it follows $P_{\rm GOE}(r)$, supporting our claim about the chaotic nature of the largest sector. We also find that the mean-gap ratio $\langle r \rangle=0.531$ is in excellent agreement with the theoretical value $\langle r \rangle_{\rm GOE}\simeq 0.536$.

To provide further evidence for weak ergodicity breaking, we computed the half-chain entanglement entropy for all eigenstates of the model in a finite volume. States in the sectors with ergodicity breaking are expected to have relatively small entanglement, whereas, for the majority of eigenstates, we expect volume-law entanglement, which in the thermodynamic limit only depends on the energy density of the state \cite{fragmentation-scars-review-2}. We used a volume $L=14$ and diagonalized the Hamiltonian in the zero-momentum sector of the integrable subspace and its complement separately. In this way, we can directly distinguish the entanglement properties of the two types of eigenstates.

\begin{figure}[h!]
    \centering
\includegraphics[width=1.\linewidth]{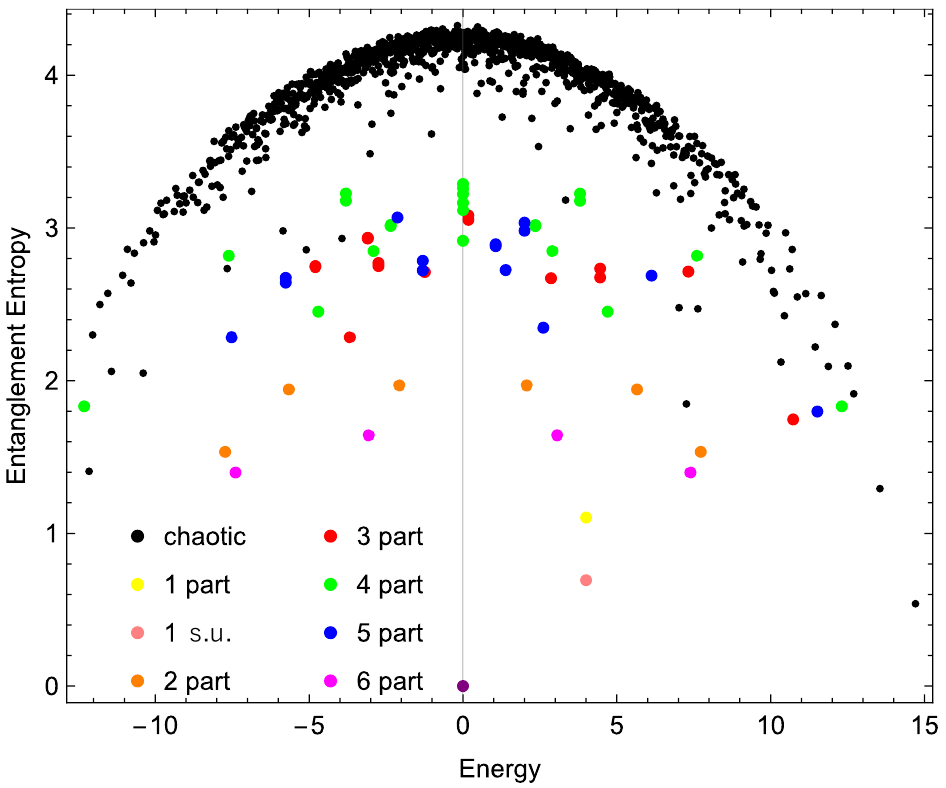}
    \caption{Plot of the half-chain entanglement entropy as a function of the energy for the zero-momentum eigenstates of the Hamiltonian \eqref{foldedpert} with $L=14$ and $\kappa=\kappa'=0.789$. {The black dots represent the states in the chaotic sector, while the colored ones correspond those in the subsectors of the integrable sector with different particles. The pink dot "1 s.u." identifies the state with one spin up.
    The purple point $(0,0)$ is degenerate and corresponds to state with $0,\,4,\,7$ particles.}   
}\label{EEEfoldedxxz}
\end{figure}

The results for the half-chain entanglement entropy are given in Fig. \ref{EEEfoldedxxz}. We find that most points from the chaotic sector 
collapse onto a curve, and these states have relatively large entanglement. 
In contrast, states in the integrable sector have lower entanglement and remain scattered, rather than clustering around a curve. 
This is characteristic of integrable models, see, e.g., \cite{exc-state-entr, rigol-entangl-xxz,swietek-entangl-xyz}. Furthermore, we have identified each state in the integrable sector and have checked that the corresponding energy can also be calculated via Eq. \eqref{solutionfromBA}, the one predicted with the Bethe ansatz. As for the adjacency graph, there is one state that does not belong to the integrable sectors with a fixed number of particles (pairs of down spin). This corresponds to a state with one spin 
up and hence is trivially solvable.

Finally, we examine how the ratio of the dimension of the chaotic sector to that of the full Hilbert space varies with the volume $L$. Let $D_{\rm chaos}(L)$ be the dimension of the chaotic sector for chain length $L$. We found that 
 $D_{\rm chaos}(L)$ follows the sequence
\begin{equation}
\label{chaos}
    D_{\rm chaos}(L) = 
    \begin{cases} 
        2^L - \mathsf{L}_L - 1 & \text{if } L \text{ is odd} \\
        2^L - \mathsf{L}_L - L + 1 & \text{if } L \text{ is even}
    \end{cases},
\end{equation}
where $\mathsf{L}_L$ is the $L$th Lucas number defined by the recurrence relation $\mathsf{L}_L=\mathsf{L}_{L-1}+\mathsf{L}_{L-2}$ with the initial conditions $\mathsf{L}_0=2$ and $\mathsf{L}_1=1$.

The above result for $D_{\rm chaos}(L)$ can be understood by computing the dimensions of the integrable subspaces. The distinguished sector that we treated above in Section \ref{sec:foldedint} has dimension 
\begin{equation}
D_{\rm constr}=    \begin{cases}
        \mathsf{L}_L & \text{if } L \text{ is odd} \\ 
       \mathsf{L}_L -1  & \text{if } L \text{ is even} \\ 
    \end{cases}
\end{equation}
If the volume is odd, then one gets an additional ``integrable'' subspace spanned by the state with all spins down, and this is not included in $D_{\rm constr}$ by the definition of that subspace. Finally, if the volume is even, one has two additional integrable subspaces spanned by the states with only one up spin that propagates on the even or the odd sites, respectively. This gives an additional dimension $L$ in the even case.  Putting everything together, one obtains the formula \eqref{chaos}. We also checked this formula up to $L=22$ by analyzing the adjacency graphs directly. 

Since $\mathsf{L}_L \sim \varphi^L$, where $\varphi=(1+\sqrt{5})/2 \sim 1.618$ is the golden ratio, the ratio between $D_{\rm chaos}(L)$ and the dimension of the full Hilbert space is given by
\begin{equation}
\frac{D_{\rm chaos}}{2^L} \sim 1- (0.809)^L,
\end{equation}
which approaches $1$ exponentially as $L$ increases. In other words, the non-ETH subspace is a vanishing fraction of the full Hilbert space. This further confirms that our model weakly breaks ergodicity (see Definition \ref{defweakergb} for the definition of weak ergodicity breaking).

\section{The dipole conserving model and its perturbations}

In this section, we treat the dipole-conserving model introduced in
\cite{tibor-fragment}. The model displays Hilbert space fragmentation, and it was already found in
\cite{tibor-fragment,tibor-fragm-SLIOM} that the model has integrable sectors.
We add perturbation terms to the model, which will restore ergodicity for almost all states, nevertheless keeping selected
integrable sectors intact.

The model is a spin-$1$ chain with a Hamiltonian with three-site interactions. The original Hamiltonian is defined as
\begin{equation}
  H_0=\sum_j h(j)
  \label{eq:dipoleHam}
\end{equation}
with
\begin{equation}
\label{eq:local_dipole_ham}
h(j)=  S_j^-(S^+_{j+1})^2S^-_{j+2}+S_j^+(S^-_{j+1})^2S^+_{j+2}.
\end{equation}
Here $S^{\pm}_j$ are the standard raising/lowering operators acting on site $j$. This Hamiltonian has no free parameters.

This model preserves the total magnetization and also the dipole moment defined by
\begin{equation}
  S^z=\sum_j S^z_j,\qquad D=\sum_j  j \,S^z_j.
\end{equation}
Let us denote the local basis states as $\ket{+}$, $\ket{0}$ and $\ket{-}$. We say that $\ket{0}$ is the vacuum state
and $\ket{+}$ and $\ket{-}$ are charges of different sign above the vacuum. Dipole conservation implies that an isolated
charge cannot move in the vacuum, unless it emits a dipole. However, dipoles on their own are mobile. The two dipole
configurations are $\ket{+-}$ and 
$\ket{-+}$, and the Hamiltonian generates hopping terms for them.
Inspection of the combination of raising/lowering operators indeed shows that we obtain the transition matrix elements
\begin{equation}
  \label{dipole1}
  \begin{split}
    \ket{0+-}\quad &\leftrightarrow\quad \ket{+-0}\\
      \ket{0-+}\quad &\leftrightarrow\quad \ket{-+0}\\
  \end{split}
\end{equation}
These can be interpreted as hopping terms for the dipoles. However, these are not the only transition matrix elements of
the Hamiltonian. The remaining processes are
\begin{equation}
    \label{dipole2}
  \begin{split}
    \ket{0+0}\quad &\leftrightarrow\quad \ket{+-+}\\
      \ket{0-0}\quad &\leftrightarrow\quad \ket{-+-}\\
  \end{split}
\end{equation}
These can be interpreted as emission/annihilation of dipoles with the isolated charges as seeds.

Fragmentation in the unperturbed model was discussed in detail in \cite{tibor-fragment,tibor-fragm-SLIOM}. Here we
provide an alternative treatment by using the terminology of the ``irreducible strings''
\cite{trimer1,trimer2,dhar-symmetries}. The irreducible strings are very closely related to the ``Statistically
Localized Integrals of Motion'' introduced in \cite{tibor-fragment,tibor-fragm-SLIOM}. Once again they are defined for
the computational basis, with a procedure we now explain.

For any basis state, let us disregard the vacuum configurations; this way we obtain a sequence
of $+$ and $-$ signs. Let us now apply the following rule to the sequence: any occurrence of the triplets $+-+$ and
$-+-$ have to be replaced by $+$ and $-$, respectively. This procedure has to be applied iteratively until one reaches
a sequence with no occurrence of the distinguished triplets. It can be seen that this final sequence is independent of
the order of the shortening steps. The final sequence consists of blocks of $+$ and $-$ signs such that
each block has two or more elements (except at the boundary). We call this sequence the irreducible string. It can be
seen that this is an invariant of the motion: the kinetic terms do not affect the original sequence, and the remaining
processes create or annihilate the triplets that need to be deleted anyways to obtain the irreducible string. 

\subsection{Integrable subspace}

Simple integrable subspaces are given by the irreducible strings that are essentially empty. In the case of open
boundary conditions, let us consider the irreducible strings $+$, $-$, $+-$ and $-+$. These strings correspond to states
in the Hilbert space where every $+$ is followed by a $-$ and vice versa, with possible insertions of the local vacuum
state 0. The four different sectors correspond to four possibilities specified by the leftmost and the rightmost
charge.

It was shown in \cite{tibor-fragm-SLIOM} that the dynamics in these four sectors is equivalent to the XX model with open
boundary conditions. The correspondence is the following. Let us perform a special type of bond-site transformation, which is meaningful specifically in these four selected sectors: we
put a $\ket{\rightarrow}$ or $\ket{\leftarrow}$ to every bond between two sites, according to whether the first charge
to the right of the bond is a $\ket{+}$ or a $\ket{-}$, respectively. An equivalent rule is that
we
put a $\ket{\rightarrow}$ or $\ket{\leftarrow}$ to every bond, according to whether the first charge
to the left is a $\ket{-}$ or a $\ket{+}$, respectively. If there are charges to the left and to the right, too, then the two rules
give the same bond variable, and if there is a charge only to the left or only to the right, then the appropriate rule
needs to be applied. As an effect, the arrows on the bonds will point from the negative charges $\ket{-}$ to the
positive charges $\ket{+}$. 

Now it can be seen that all the transition matrix elements in \eqref{dipole1}-\eqref{dipole2} are translated to the
transitions
\begin{equation}
\ket{\leftarrow\rightarrow}\quad\leftrightarrow\quad\ket{\rightarrow\leftarrow}
\end{equation}
on the bond variables. These transitions define the XX model.

If we make the identifications $\ket{0}\to\ket{0}$, $\ket{+}\to\ket{1}$, $\ket{-}\to\ket{2}$, then these distinguished subspaces
become identical to the integrable subspace $\tilde \HH$ of the perturbed XXC models defined in Sec.
\ref{sec:nonintxxc}. However, the mechanism of Hilbert space fragmentation differs in the two models, and also the definition of
the irreducible string differs. What is more, the specific embedding of the XX model is also different. It is only the
constrained sectors themselves that are identical in these two models.

\subsection{Weak ergodicity breaking}

{
Now we introduce a perturbed Hamiltonian as
\begin{equation}
  \label{dipoleX}
    H=\sum_j h(j)+h^{2{\rm f}} (j) + h^{3{\rm f}} (j), 
\end{equation}
where 
$h(j)$ is defined in Eq. \eqref{eq:local_dipole_ham}, 
\begin{equation}
  h^{2{\rm f}} (j)=\kappa
\left[    (S^-_j S^-_{j+1})^2+ (S^+_j S^+_{j+1})^2\right],
\label{eq:hX1}
\end{equation}
and 
\begin{equation}
  h^{3{\rm f}} (j) = \kappa'
\left[    (S^-_j S^-_{j+1} S^-_{j+2})^2 + {\rm H.c.}
\right].
\label{eq:hX2}
\end{equation}
It can be seen that these special perturbations keep the integrable sectors intact. This happens because the perturbation
term $h^{2{\rm f}} (j)$ in \eqref{eq:hX1} only acts on local configurations $\ket{++}$ and $\ket{--}$, and these do not occur in the integrable sectors. The term $h^{3{\rm f}} (j)$ in \eqref{eq:hX2} further introduces matrix elements between disconnected subspaces of the unperturbed Hamiltonian $H_0$, but still preserves the integrable sectors.
}

\begin{figure}
    \centering
\includegraphics[width=\linewidth]{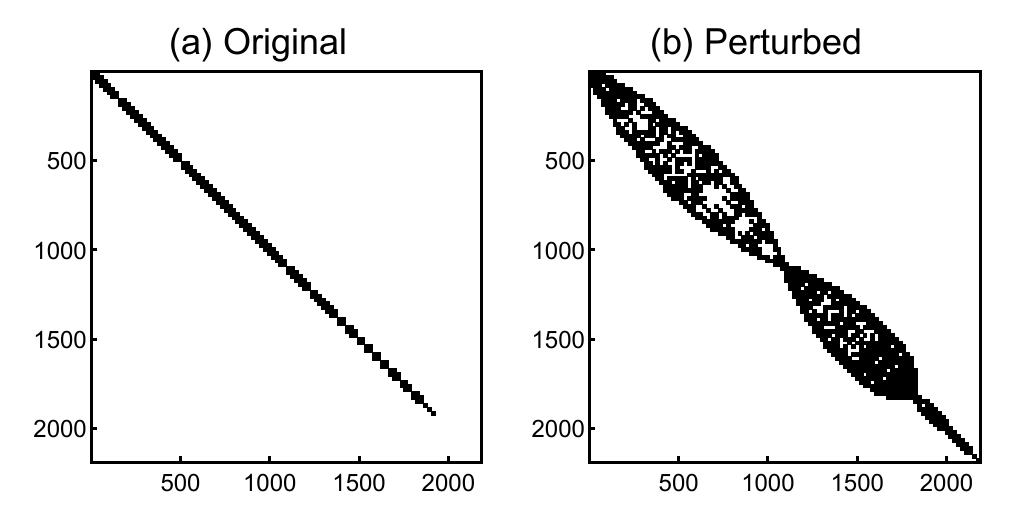}
\caption{Hamiltonian structure of (a) the original model $H_0$ in Eq. \eqref{eq:dipoleHam} and (b) the perturbed model $H$ in Eq. \eqref{dipoleX} with $\kappa=\kappa'=1.0$ for $L=7$. Black (white) pixels represent nonzero (vanishing) matrix elements. The rows and columns are downsampled so that the structures are visible in the plots.}
\label{fig:MatrixPlot_diple_L7}
\end{figure}

\begin{figure}
    \centering
\includegraphics[width=0.9 \linewidth]{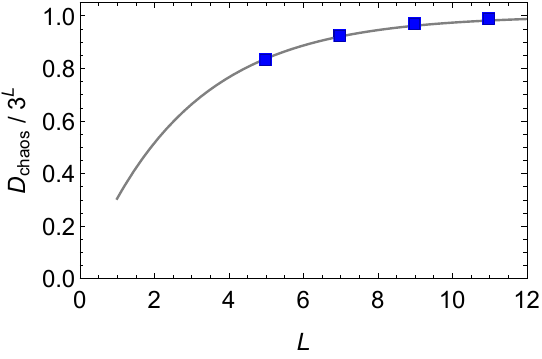}
    \caption{Ratio of the dimension of the chaotic sector to that of the full Hilbert space as a function of 
    $L$ for the Hamiltonian $H$ in Eq. \eqref{dipoleX} with $\kappa=\kappa'=1.0$. 
    Here $D_{\rm chaos}$ denotes the sum of the dimensions of the $(L-1)/2$ largest subspaces. The gray solid curve is a fit to the date using $1-\exp (-\alpha L)$ with $\alpha \simeq 0.362$. The ratio approaches $1$ as the length increases.} 
\label{fig:dim_chaos_dipole}
\end{figure}

\begin{figure}
\vspace{0.5cm}
    \centering
\includegraphics[width=0.9\linewidth]{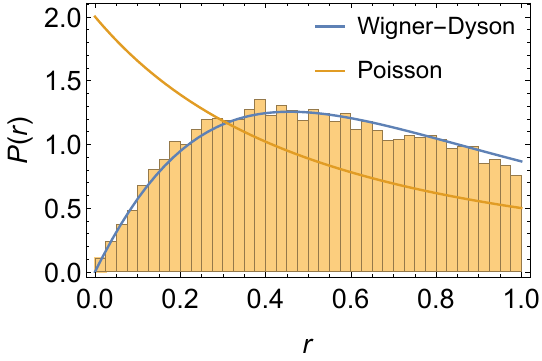}
    \caption{Distribution of the adjacent-gap ratios of the Hamiltonian $H$ in Eq. \eqref{dipoleX} for $L=13$ with $\kappa=\kappa'=1.0$. The data are collected from the subspace of the largest chaotic sector, which is invariant under translation, spin-flip ($S^z_j \to -S^z_j$), and spatial inversion. The dimension of this subspace is $15695$.  
    The mean gap ratio is $\langle r \rangle=0.526$, which is in rough agreement with $\langle r \rangle_{\rm GOE}\simeq 0.536$. 
    The GOE Wigner-Dyson and Poisson distributions are shown for comparison.} 
\label{fig:dipole_L13_level}
\end{figure}

{
To demonstrate that the perturbations introduced undo the fragmentation, we numerically study the model in a finite volume with periodic boundary conditions. We show in Fig. \ref{fig:MatrixPlot_diple_L7} the Hamiltonian structure of $H_0$ in Eq. \eqref{eq:dipoleHam} and $H$ in Eq. \eqref{dipoleX} in a properly ordered basis. Clearly, the perturbations connect exponentially many disconnected subspaces of the original model, resulting in a smaller number of sectors. Note that, in addition to the integrable sectors, there still remain many small subspaces even after the perturbations are introduced. This may be related to the conservation of the local dipole moment in the original model, which is defined within the region between two subsequent defects \cite{tibor-fragm-SLIOM}. 
}

{
While the remaining fragmentation deserves further investigation, we have found numerically that the first few largest subspaces dominate the full Hilbert space; the ratio of the sum of their dimensions and the total Hilbert space dimension approaches $1$ as $L$ increases, as shown in Fig. \ref{fig:dim_chaos_dipole}. This, together with the analysis of the adjacent-gap ratios (see Fig. \ref{fig:dipole_L13_level}), suggests that the perturbed model exhibits weak ergodicity breaking according to Definition \ref{defweakergb}.  
}

\section{Integrable models designed for constrained Hilbert spaces}

\label{sec:constr}

In the literature, there exist models that are defined on certain constrained subspaces of the configuration
space, and there is no guarantee that the Hamiltonians would be integrable when extended to the full Hilbert space. 
It is a natural idea to take these integrable models and consider their action on the full Hilbert space.
If the extension is such that the constrained Hilbert space is preserved, then we can find a non-integrable model with an isolated integrable sector. This process typically results in strong ergodicity breaking via fragmentation. However, this fragmentation can be undone by adding the appropriate perturbations, thereby providing a somewhat different mechanism for finding an isolated integrable sector in a chaotic model.

In two-dimensional (2D) statistical mechanics models, famous examples for constrained models are the so-called restricted solid-on-solid (RSOS) models, 
whose study goes back to Baxter \cite{Baxter-Book}. These models are in direct correspondence with spin chain
Hamiltonians, which act on a constrained Hilbert space, such that the constraint follows directly from the restrictions
of the 2D statistical mechanics model.

Perhaps the most important example for such Hamiltonians is the RSOS
family \cite{RSOS-H}. 
In this case, one deals with the Rydberg constraint: the restricted Hilbert space 
consists of states in the computational basis 
that do not have two down spins on neighboring sites. 
A family of non-integrable Hamiltonians acting on this restricted space was studied much earlier in
\cite{fendley-sachdev-rsos, lesanovsky2012interacting}. This family of model Hamiltonians has two free parameters, and after a specialization one
obtains the one-parameter family of integrable models derived, for example, in \cite{RSOS-H}. This family includes the
so-called golden chain as a special point \cite{golden-chain}; this critical Hamiltonian is related to the fusion rules of anyons and also
to the Temperley-Lieb algebra. Other integrable models defined on this restricted Hilbert space were constructed
recently in \cite{sajat-rydberg} (for interesting developments on Hilbert spaces with very different constraints, see
\cite{marius-haagerup}). 

If we take such an integrable Hamiltonian that acts on the restricted Hilbert space and extend it formally so
that it should act on the full Hilbert space, one typically ends up with immobile local configurations, which in turn leads to
strong Hilbert space fragmentation. For an earlier discussion of this phenomenon in a related model, see, for example,
\cite{rule201}.

In order to circumvent this problem and obtain a thermalizing Hamiltonian for the complementary part of the Hilbert
space, we add perturbation terms that create hopping processes for otherwise immobile local configurations, 
while keeping the constrained subspace intact. This way, we obtain a new mechanism for weak ergodicity breaking.

A crucial difference between this approach and our other examples is that now the integrable model for the constrained subspace does not necessarily have any kind of particle conservation.
To illustrate this, we now consider a specific example of constrained
Hilbert spaces, namely the Rydberg constraint.

\subsection{The spin-$1/2$ off-critical RSOS chains}

In this case, the constraint is as follows: in the computational basis, we select those states where 
no two down spins 
are adjacent. This constraint naturally arises in experiments with Rydberg atoms
\cite{rydberg-blockade-experimentally}, and is often called Rydberg blockade or Rydberg constraint.
Certain non-integrable models defined for this Hilbert space (namely the so-called PXP model and its generalizations)
host quantum many-body scars \cite{scars-elso, PXP-2, hfrag-review}. 

Integrable Hamiltonians acting on this constrained Hilbert space, and having three-site interactions were 
studied in
\cite{RSOS-H}, although the crucial elements of the theory go back to \cite{RSOS-1,RSOS-2}. These integrable
Hamiltonians are given by
\begin{equation}
  H^{\rm int}=\sum_j h^{\rm int}(j),
  \label{eq:int_RSOS}
\end{equation}
where 
\begin{equation}
  h^{\rm int}(j)=P_j X_{j+1}P_{j+2} +a N_{j+1}+b N_jN_{j+2}
\end{equation}
with
\begin{equation}
  \label{condition}
  ab+b^2 = 1,
\end{equation}
and periodic boundary conditions are imposed.
Notice that this Hamiltonian does not preserve the magnetization due to the single spin-flip operator $X_{j+1}$.

Let us consider the action of this Hamiltonian on the full Hilbert space. In this case, the integrability of the full model is strictly speaking lost. We find that two neighboring down spins
become fully immobile under the action of this Hamiltonian. This implies that the full Hilbert space fragments into
exponentially many sectors according to the placements of blocks of down spins.

In order to circumvent this fragmentation, we consider the following Hamiltonian acting {\it on the full} Hilbert space
\begin{equation}
    H=\sum_j h^{\rm int}(j)+\kappa \, h^{\rm kin}(j)+\kappa' h^{X}(j).
\label{eq:deformedRSOS}
\end{equation}
Here $h^{\rm kin}(j)$ is given by \eqref{foldedb} and
\begin{equation}
    h^{X}(j)=X_jN_{j+1}N_{j+2}+N_j N_{j+1}X_{j+2}.
\end{equation}
It is important that both $h^{\rm kin}(j)$ and $h^{X}(j)$ act as identically zero on the constrained Hilbert space,
therefore, they preserve it. However, they undo the fragmentation. The operator $h^{\rm kin}(j)$ provides a kinetic term for
a block of two neighboring down spins, and $h^{X}(j)$ further changes the number of down spins, but only if there are
two or more such down spins at neighboring positions.

We claim that these perturbations undo the fragmentation such that the full model becomes ergodic. This claim is justified below by numerical computations.

\begin{figure}[h!]
    \centering
    \includegraphics[width=0.85\linewidth]{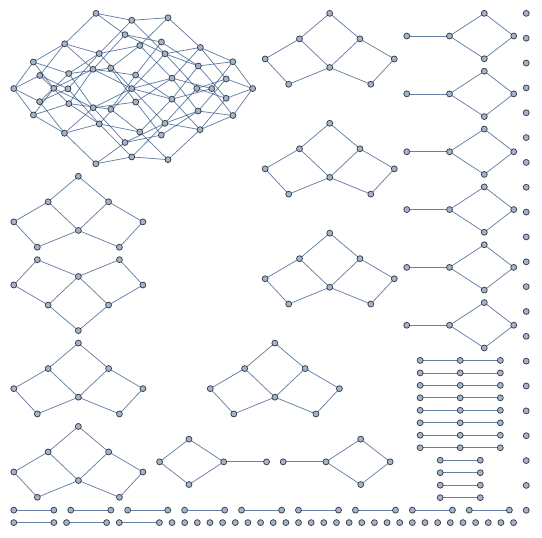}
    \caption{Adjacency graph of the Hamiltonian $H$ in Eq. \eqref{eq:deformedRSOS} for $L=8$ with $a=1.5$, $b=0.5$, and $\kappa=\kappa'=0$. {The upper left sector corresponds to the integrable model in the constrained Hilbert space. For better visibility, we have removed the loops in the graph.}}
    \label{unperturbed_RSOS_connectivity_L8}
\end{figure}

\begin{figure}[h!]
    \centering
    \includegraphics[width=1.0\linewidth]{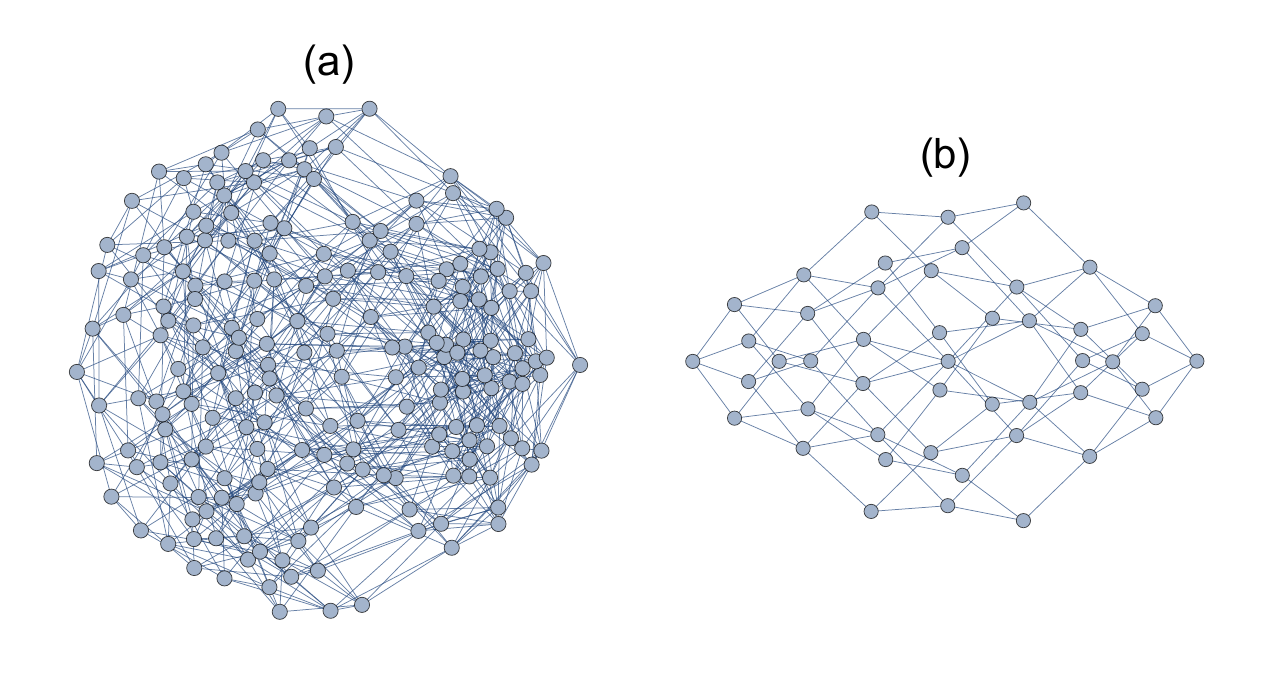}
    \caption{Adjacency graph of the Hamiltonian $H$ in Eq. \eqref{eq:deformedRSOS} for $L=8$ with $a=1.5$, $b=0.5$, and $\kappa=\kappa'=1.0$. (a) Chaotic sector with $209$ states. (b) Integrable sector with $47$ states,  
    corresponding to the upper left sector in the adjacency graph of the undeformed model shown in Fig. \ref{unperturbed_RSOS_connectivity_L8}.}
    \label{RSOS_connectivity_L8}
\end{figure}

\subsection{Numerical investigations}
As before, we demonstrate numerically that the Hamiltonian $H$ in Eq. \eqref{eq:deformedRSOS} describes a model with weak ergodicity breaking. In this model, the integrability is manifest in the constrained Hilbert space, where two down spins are not allowed to occupy 
adjacent sites. In the full Hilbert space, the model exhibits fragmentation. We show that after adding the two engineered perturbations, the integrable sector is preserved, but in the complementary space, the remaining sectors are not fragmented anymore and are merged into a chaotic sector.

\begin{figure}
\vspace{0.5cm}
    \centering
\includegraphics[width=0.9\linewidth]{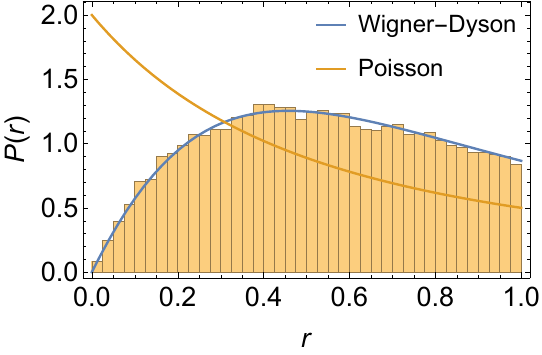}
    \caption{Distribution of the adjacent-gap ratios of the Hamiltonian $H$ in Eq. \eqref{eq:deformedRSOS} for $L=20$ with $a=1.5$, $b=0.5$, and $\kappa=\kappa'=1.0$. The data are collected from the subspace of the chaotic sector, which is translationally invariant and even under spatial inversion. The dimension of this subspace is $26557$. 
    The GOE Wigner-Dyson and Poisson distributions are shown for comparison.} \label{RSOS_L20_level}
\end{figure}

\begin{figure}

\vspace{0.5cm}
    \centering
\includegraphics[width=0.9\linewidth]{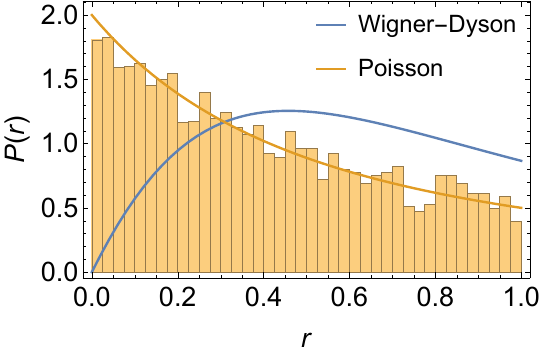}
    \caption{Distribution of the adjacent-gap ratios of the Hamiltonian $H$ in Eq. \eqref{eq:deformedRSOS} for $L=25$ with $a=1.5$, $b=0.5$, and $\kappa=\kappa'=1.0$. The data are collected from the subspace of the integrable sector, which is translationally invariant and even under spatial inversion. The dimension of this subspace is $3545$. 
    The GOE Wigner-Dyson and Poisson distributions are shown for comparison.} 
\label{RSOS_L25_integrable_level}
\end{figure}

First, we show in Fig. \ref{unperturbed_RSOS_connectivity_L8} 
the adjacency graph for the unperturbed Hamiltonian  
highlighting the presence of many subsectors. 
This fragmentation arises from the fact that  
a block of two adjacent down spins is immobile. 
Then, we add the perturbation terms $h^{\rm kin}(j)$ and $h^X(j)$ that undo the fragmentation. 
We show the adjacency graph of the perturbed model in Fig. \ref{RSOS_connectivity_L8}, where only two sectors are present. We claim that the larger sector (a) corresponds to the chaotic sector. To confirm this, we show in Fig. \ref{RSOS_L20_level} that the adjacent-gap ratio distribution for sector (a) follows the Wigner-Dyson distribution. The other sector, (b) in Fig. \ref{RSOS_connectivity_L8}, is the integrable sector, 
which can be identified in the adjacency graph of the unperturbed model shown in Fig. \ref{unperturbed_RSOS_connectivity_L8}. The adjacent-gap ratio distribution for this sector  
follows the Poisson distribution, as shown in Fig \ref{RSOS_L25_integrable_level}. 
We calculate the mean-gap ratios and obtain $\langle r \rangle=0.532$ for sector (a) and $\langle r \rangle = 0.394$ for sector (b), which supporting our claim. 

\begin{figure}
    \centering
    \includegraphics[width=0.8\linewidth]{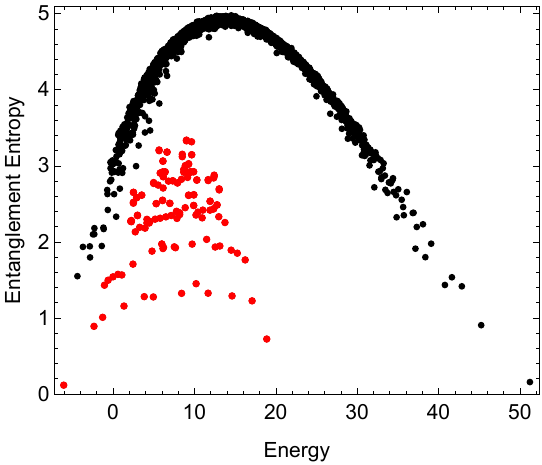}
    \caption{Plot of energy vs half-chain entanglement entropy for the zero-momentum eigenstates of the deformed RSOS model \eqref{eq:deformedRSOS} with $L=16$, $a=1.5$, $b=0.5$, and $\kappa=\kappa'=1$. The black dots represent the states in the chaotic sector, while the red ones correspond to those in the integrable sector.}
    \label{RSOS_L16_EE}
\end{figure}

Next, we numerically compute the half-chain entanglement entropy, which is an indicator of weak ergodicity breaking. Figure \ref{RSOS_L16_EE} shows the results for the zero-momentum sector of the Hamiltonian $H$ in Eq. \eqref{eq:deformedRSOS} with $L=16$, $a=1.5$, $b=0.5$, and $\kappa=\kappa'=1$. We observe that the states in the chaotic sector collapse onto a curve and have higher entropy compared to those in the integrable sector. Interestingly, the low-entanglement states appear to form multiple arcs, reminiscent of quantum many-body scars in the PXP model \cite{scars-elso, PXP-2}. However, elucidating the underlying mechanism is beyond the scope of this work. 

Finally, we discuss the ratio of the dimension of the chaotic sector to that of the full Hilbert space. It is well known that the dimension of the restricted Hilbert space subject to the Rydberg constraint is equal to the $L$th Lucas number for a chain of length $L$ with periodic boundary conditions \cite{scars-elso}. Consequently, the dimension $D_{\rm chaos}(L)$ of the chaotic sector is given by $D_{\rm chaos}(L)=2^L-{\sf L}_L$, where ${\sf L}_L$ is the $L$th Lucas number, as before. This leads to $D_{\rm chaos}/2^L \sim 1-(0.809)^L$, which approaches $1$ exponentially with $L$, further evidencing weak ergodicity breaking in our model.


\section{Quantum circuits}
\label{quantumcircuits}
In this section, we show that the embeddings of integrable models are not restricted to continuous-time evolution, but
also to certain types of quantum circuits that describe discrete-time evolution. It is known that for integrable models
associated with the Yang-Baxter equation there exist integrable Trotterizations, which are regular quantum circuits that
approximate continuous-time evolution \cite{integrable-trotterization}. Here we show that the Hamiltonians described
above can be approximated by similar quantum circuits such that there is a distinguished integrable sector where
dynamics coincides with that of the corresponding integrable Trotterization.

The idea is to take a model Hamiltonian
\begin{equation}
  H=H^{\rm int}+H^X=\sum_j   h^{\rm int}(j)+h^X(j),
\end{equation}
where $h^{\rm int}(j)$ is the integrable part, and $h^X(j)$ is a perturbation. We approximate the continuous time evolution
as
\begin{equation}
  e^{-iH t}\approx (U^{\rm int}U^X)^N,
\end{equation}
where $U^{\rm int}$ and $U^X$ are unitary operators constructed using local quantum gates, and $N$ is the so-called Trotter
number. It follows that the product $U^{\rm int}U^X$ is assumed to approximate $e^{-iH (\Delta t)}$, where $\Delta t=t/N$. 

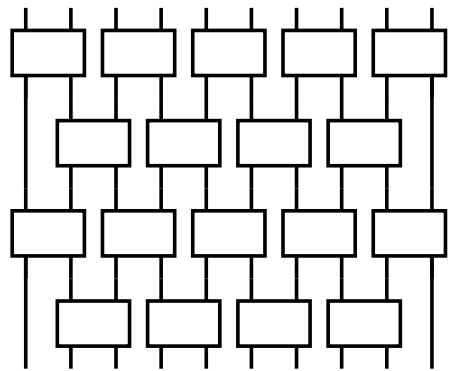
\begin{figure}
    \centering
  
\tikzset{every picture/.style={line width=0.8mm}}
\scalebox{0.6}{
\begin{tikzpicture}

  \draw  (0,0) -- (0,-2.5);
  \draw  (9,0) -- (9,-2.5);

    \draw (0,4) -- (0,1.5);
    \draw  (9,4) -- (9,1.5);

  \foreach \j in {0,2,4,6,8}
  {
    \begin{scope}[xshift=\j cm]
  \draw (-0.3,0) rectangle (1.3,1);
\draw  (0,1.5) -- (0,1);
\draw  (1,1.5) -- (1,1);
\draw(0,0) -- (0,-0.5);
\draw  (1,0) -- (1,-0.5);
\end{scope} 
}

\begin{scope}[yshift=4 cm]
    \foreach \j in {0,2,4,6,8}
  {
    \begin{scope}[xshift=\j cm]
  \draw (-0.3,0) rectangle (1.3,1);
\draw  (0,1.5) -- (0,1);
\draw  (1,1.5) -- (1,1);
\draw  (0,0) -- (0,-0.5);
\draw  (1,0) -- (1,-0.5);

\end{scope} 
}
\end{scope}

\begin{scope}[yshift=2cm]
    \foreach \j in {1,3,5,7}
  {
    \begin{scope}[xshift=\j cm]
  \draw (-0.3,0) rectangle (1.3,1);
\draw  (0,1.5) -- (0,1);
\draw  (1,1.5) -- (1,1);
\draw  (0,0) -- (0,-0.5);
\draw  (1,0) -- (1,-0.5);
\end{scope} 
}
\end{scope}

\begin{scope}[yshift=-2cm]
    \foreach \j in {1,3,5,7}
  {
    \begin{scope}[xshift=\j cm]
  \draw (-0.3,0) rectangle (1.3,1);
\draw  (0,1.5) -- (0,1);
\draw  (1,1.5) -- (1,1);
\draw    (0,0) -- (0,-0.5);
\draw    (1,0) -- (1,-0.5);
\end{scope} 
}
\end{scope}

\end{tikzpicture}}
    
    \caption{An example for a brickwork circuit with two-site gates and  open boundary condition. It is understood that time flows in the vertical direction. The wires stand for the qudits and boxes denote two-site gates. }
    \label{fig:brick1}
\end{figure}

We will construct circuits such that
\begin{equation}
  U^{\rm int}\approx e^{-iH^{\rm int} (\Delta t)},\qquad U^{X}\approx e^{-iH^{X} (\Delta t)}
\end{equation}
We will use the so-called brickwork circuits for both unitary operators; for a depiction of a brickwork circuit see Fig. \ref{fig:brick1}. The key idea is that $U^{\rm int}$ will be given by
exactly the same brickwork circuit as for the purely integrable model, whereas $U^X$ will be given by a brickwork
circuit that uses only the perturbation terms. This implies that the integrable sectors will be the same as in the case
of the Hamiltonians. The actual time evolution will be slightly different due to the discretization, but the kinetic constraints
and the conserved quantities (irreducible strings) will not be affected by the discretization. The fact that the continuous
and discrete time evolution can lead to the same Hilbert space fragmentation was already noticed in
\cite{tibor-fragment,fragment-fracton-2}.

The concrete formulas for $U^{\rm int}$ and $U^X$  depend on the model. The circuit geometry depends slightly on the
interaction range of the Hamiltonian. Nearest neighbor interacting models are the simplest. In those cases one can
apply the formulas of \cite{integrable-trotterization}. In the case of three-site and four-site interactions, the integrable Trotterizations can be constructed using the techniques of \cite{sajat-medium}.

In this section, we provide examples for these discretizations, but we do not give the final formulas for every embedding
that we considered in the previous sections. Here our goal is simply to demonstrate that the integrable embeddings can
be lifted also to the circuits.

\subsection{Circuits for the perturbed Maassarani-Mathieu spin chain}

Now we consider the model defined in Sec. \ref{sec:MM}. More specifically we focus on the Hamiltonian
\begin{equation}
  \label{XXCH3}
    H=\sum_j   h^{\rm kin}_{j,j+1}+h^{\rm pair}_{j,j+1},
\end{equation}
where $h^{\rm kin}_{j,j+1}$ and $h^{\rm pair}_{j,j+1}$ are given by \eqref{xxckin} and \eqref{gammatermXXX}, respectively. We
set the diagonal interaction terms to zero for simplicity, such that the kinetic term describes the original
Maassarani-Mathieu spin chain \cite{su3-xx}.

Now we discretize the time evolution dictated by this Hamiltonian. We construct a quantum circuit of the form
\begin{equation}
  U=U^{\rm int}U^X,
\end{equation}
where both factors are given by a ``brickwork'' geometry:
\begin{equation}
  \begin{split}
    U^{\rm int}&=\prod_{j=1}^{L/2} V^{\rm int}_{2j,2j+1}\prod_{j=1}^{L/2} V^{\rm int}_{2j-1,2j}\\
    U^{X}&=\prod_{j=1}^{L/2} V^{X}_{2j,2j+1}\prod_{j=1}^{L/2} V^{X}_{2j-1,2j},\\
    \end{split}  
\end{equation}
where $V^{\rm int}_{k,k+1}$ and $V^{X}_{k,k+1}$ are two-site gates. For the integrable part we choose the gate to be an
evaluation of the so-called $\check R$ matrix \cite{integrable-trotterization}:
\begin{equation}
  V^{\rm int}_{k,k+1}=\check R(-\mu),
\end{equation}
where $\mu\in\valos$ is a parameter of the circuit, and the $\check R$ matrix is the solution of the Yang-Baxter relation
corresponding to this model \cite{su3-xx}. It is given by \cite{su3-xx}
\begin{equation}
  \begin{split}
  \check R_{j,k}(\mu)=&i\sin(\mu) h^{\rm kin}_{j,k}+
 (N^{\rm tot}_jP_k+P_j N^{\rm tot}_k)+\\
& + \cos(\mu)(P_jP_k+N^{\rm tot}_jN^{\rm tot}_k),
  \end{split}
\end{equation}
where we used the notations introduced in Sec. \ref{sec:trivial}. 
For small values of $\mu$, the operator $\check
R_{j,k}(\mu)$ is a good approximation for $\exp(-ih^{\rm kin}_{j,j+1}\mu)$.

It follows from the general construction in \cite{integrable-trotterization} that time evolution by $U^{\rm int}$ only gives
an integrable Trotterization for the Maassarani-Mathieu spin chain.

For the circuit responsible for the perturbation we can simply choose
\begin{equation}
  V^{X}_{k,k+1}=\exp (-ih^{\rm pair}_{k,k+1}\mu).
\end{equation}
The introduction of these two-site gates generally breaks integrability. Nevertheless, these two-site gates act as 
the identity on the distinguished integrable subspace, therefore they preserve the integrability of the distinguished sector.

\subsection{Circuits for the perturbed folded XXZ model}

Now we consider the perturbed model treated in Sec. \ref{sec:foldedpert}. More concretely, we intend to discretize
the time evolution given by the Hamiltonian
\begin{equation}
  H=H^{\rm int}+H^X,
\end{equation}
where now
\begin{equation}
  H^{\rm int}=\sum_j h^{\rm kin}(j),
\end{equation}
where $h^{\rm kin}(j)$ is given by \eqref{foldedb}, and
\begin{equation}
    H^X=\sum_{j} \tilde h^{\rm kin}(j)+h^X(j)
\end{equation}
where $\tilde h^{\rm kin}(j)$ and $h^X(j)$ are given by eqs. \eqref{foldedx1}-\eqref{foldedx2}.

The integrable part of the Hamiltonian can be discretized by the constructions in \cite{sajat-medium}, using the Lax
operators given in \cite{sajat-hardrod}. This construction is a generalization of the usual algebraic treatment of
integrable spin chains, by extending the formalism to the so-called medium range interactions, which span a few sites
instead of just two. Here we do not discuss this construction, we just refer the reader to the papers
\cite{sajat-medium,sajat-hardrod}. However, we give the concrete formulas for the circuits.

For technical reasons discussed below, let us assume that $L$ is a multiple of $3$. 
Then the integrable part is discretized as
\begin{equation}
  V^{\rm int}=\prod_{k=1,2,3}  \prod_{j=0}^{L/3-1} V_{3j+k,3j+k+1,3j+k+2},
\end{equation}
where $V_{j,j+1,j+2}$ is given by an evaluation of the so-called Lax operator, more concretely
\begin{equation}
  V_{j,j+1,j+2}=P_{j+1} \check R_{j,j+2}(-\mu),
\end{equation}
where $\mu$ is a real parameter, and $\check R$ is the $R$-matrix of the XX model acting on two sites, given concretely
by
\begin{equation}
  \check R_{j,k}(-\mu)=
  \begin{pmatrix}
    1 & & & \\
    & \cos(\mu) & -i\sin(\mu) & \\
    & -i\sin(\mu) & \cos(\mu) & \\
    & & & 1
  \end{pmatrix}.
\end{equation}

For the non-integrable part we can choose multiple types of discretizations. The perturbations have 4 site interactions,
so the most homogeneous discretization can be done if $L$ is also a multiple of 4, and we build
\begin{equation}
  V^X=\prod_{k=1,2,3,4}  \prod_{j=0}^{L/4-1} V^X_{4j+k,4j+k+1,4j+k+2,4j+k+3},
\end{equation}
where now
\begin{equation}
  V^X_{j,j+1,j+2,j+3}=\exp [ {-i\mu( \tilde h^{\rm kin}(j)+h^X(j))} ].
\end{equation}

\section{Conclusions}

\label{sec:concl}

In this work, we uncovered multiple mechanisms that allow for the embedding of an integrable model into the Hilbert space
of an otherwise ergodic model. A distinguished property of the embeddings was that the integrable subspace is not of product form with respect to real space. Alternatively, the projector to the integrable subspace has non-zero operator
entanglement.

Interestingly, we found that for all such nontrivial embeddings the models are perturbations of simpler models with
Hilbert space fragmentation. The perturbations are such that they preserve the distinguished integrable subspace, but
they connect all (or almost all) other subspaces, making the model ergodic in the complement of the integrable
subspace. The preservation of the integrable subspace appears to be connected to conservation laws of the fragmented
models. These observations hold for all our examples, including the model of \cite{znidaric-coexistence}, which appears
to be the only known example in the literature for the special type of ergodicity breaking that we consider.

It is an interesting question whether there are other mechanisms for non-trivial embeddings of integrable models. In our
examples, the distinguished subspaces could be identified by certain spin patterns, and the basis for the subspace could
be easily prepared in the computational basis. It is an interesting question whether there are more complicated rules
for the integrable subspace, such that the model Hamiltonians remain local.

In our examples, we managed to embed the XX model, the XXZ model, the constrained XXZ model, and the so-called integrable
RSOS chains into the Hilbert spaces of chaotic models. Interestingly, we did not find any mechanism that could embed the
XYZ chain. This remains an open problem for future works.

Finally, we remark that some of our examples can be used to construct models hosting quantum many-body scars. The key idea is to introduce perturbations that annihilate the target scar states while breaking the integrability of the distinguished subspaces. For instance, following the method in \cite{onsager-scars}, the XX model embedded in an integrable subspace can be perturbed such that the resulting Hamiltonian leaves a set of exact eigenstates intact.

\vspace{1cm}
{\bf Acknowledgments} 
H.K. was supported by JSPS KAKENHI Grants No. JP23K25783, No. JP23K25790, and MEXT KAKENHI Grant-in-Aid for Transformative Research Areas A “Extreme Universe” (KAKENHI Grant No. JP21H05191). C.M. was supported by JSPS KAKENHI Grant No. JP23K03244. C.P. was supported by funds from the European Union HORIZON-CL4-2022-QUANTUM-02 SGA through PASQuanS2.1 (Grant Agreement No. 101113690). B.P. was supported by the
Hungarian National Research, Development and Innovation Office, NKFIH Grant No. K-145904 and the 
NKFIH excellence grant TKP2021-NKTA-64.

\bigskip


%

\end{document}